\documentclass[letter,12pt]{article}
\usepackage{amsmath,amssymb,amsthm,mathtools}
\usepackage{natbib}
\usepackage{graphicx}
\usepackage[nodisplayskipstretch]{setspace} %
\usepackage{enumitem}
\usepackage{fullpage,rotating,afterpage}
\usepackage{microtype} %

\usepackage[T1]{fontenc} 
\usepackage[utf8]{inputenc} 

\usepackage{wrapfig,boxedminipage}

\usepackage{tikz}

\usepackage{pdflscape}
\usepackage{rotating}
\usepackage{booktabs}
\usepackage[justification=justified]{caption}
\usepackage[flushleft]{threeparttable}
\usepackage{multirow}
\usepackage{rotating}

\usepackage{scalerel,stackengine}
\stackMath
\newcommand\reallywidecheck[1]{%
	\savestack{\tmpbox}{\stretchto{%
			\scaleto{%
				\scalerel*[\widthof{\ensuremath{#1}}]{\kern-.6pt\bigwedge\kern-.6pt}%
				{\rule[-\textheight/2]{1ex}{\textheight}}
			}{\textheight}%
		}{0.5ex}}%
	\stackon[1pt]{#1}{\scalebox{-1}{\tmpbox}}%
}

\newcommand{\E}{\mathbb{E}}
\newcommand{\V}{\mathbb{V}}
\newcommand{\IR}{\mathbb{R}}

\newcommand{\IP}{\mathbb{P}}

\newcommand{\1}[1]{\mathbf{1}\{#1\}}

\newcommand{\Cov}{\textrm{Cov}}
\newcommand{\supp}{\textrm{supp}}

\newcommand{\sgn}{\textrm{sign}}
\newtheorem{proposition}{Proposition}

\newcommand{\hopt}{h_M}
\newcommand{\hhopt}{\widehat{h}_M}

\newcommand{\poi}{\theta}
\newcommand{\hpoi}{\widehat\theta}

\newcommand{\htauT}{\widehat{\tau}_T}
\newcommand{\htauY}{\widehat{\tau}_Y}

\newcommand{\h}{h_M}

\newcommand{\Xn}{\mathcal{X}_n}
\newcommand{\cv}{\textnormal{cv}_{1-\alpha}}

\numberwithin{equation}{section}

\usepackage{verbatim,color}

\def\boxit#1{\vbox{\hrule\hbox{\vrule\kern6pt
          \vbox{\kern6pt#1\kern6pt}\kern6pt\vrule}\hrule}}

\DeclareMathOperator*{\argmin}{argmin}

\newcommand{\captionfonts}{\small}

\makeatletter  
\long\def\@makecaption#1#2{%
  \vskip\abovecaptionskip
  \sbox\@tempboxa{{\captionfonts #1: #2}}%
  \ifdim \wd\@tempboxa >\hsize
    {\captionfonts #1: #2\par}
  \else
    \hbox to\hsize{\hfil\box\@tempboxa\hfil}%
  \fi
  \vskip\belowcaptionskip}
\makeatother   

 \usepackage{titlesec}
 \titleformat{\section}[block]{\centering\normalfont}{\thesection.}{0.5em plus .1em minus .1em}{\uppercase }
 \titleformat{\subsection}[runin]{\normalfont}{\thesubsection.}{0.3em plus .1em minus .1em}{\bfseries}[.]
 \titleformat{\subsubsection}[runin]{\normalfont}{\thesubsubsection.}{0.3em plus .1em minus .1em}{\it}[.]
 \titlespacing*\section{0pt}{14pt plus 2pt minus 2pt}{2pt plus 1pt minus 1pt}
 \titlespacing*\subsection{0pt}{7pt plus 2pt minus 2pt}{5pt plus 1pt minus 1pt }
 \titlespacing*\subsubsection{0pt}{4pt plus 1pt minus 1pt}{5pt plus 1pt minus 1pt}
\usepackage{chngcntr}
\usepackage{apptools}
\AtAppendix{\counterwithin{lemma}{section}}
 \AtAppendix{\counterwithin{assumption}{section}}
 \AtAppendix{\counterwithin{corollary}{section}}
  \AtAppendix{\counterwithin{theorem}{section}}

 \makeatletter
 \def\mythanks#1{%
 	\protected@xdef \@thanks {\@thanks \protect \footnotetext [\the \c@footnote ]{#1}}%
 }
 \makeatother

\title{\bfseries \large\uppercase{Bias-Aware Inference in Fuzzy Regression Discontinuity Designs}\mythanks{First Version: June 11, 2019. This Version: \today.
We would like to thank, Tim Armstrong, Marinho Bertanha, Yingying Dong, Keisuke Hirano,  Michal Koles{\'a}r,  the anonymous referees  and numerous seminar participants for  their helpful comments and suggestions.
The authors gratefully acknowledge financial support by the European Research Council (ERC) through grant SH1-77202. Contact information: 
Claudia Noack, Nuffield College and Department of Economics, University of Oxford, email: claudia.noack@economics.ox.ac.uk, website: http://claudianoack.github.io. 
Christoph Rothe, Department of Economics, University of Mannheim, 68131 Mannheim, Germany,
email: rothe@vwl.uni-mannheim.de, website: http://www.christophrothe.net.  }}

\author{\textsc{Claudia Noack}  \and \textsc{Christoph Rothe}}
\date{\vspace{-1cm}}

\usepackage{xr}
\begin{document}

\pagestyle{plain}

\newtheorem{theorem}{Theorem}
\newtheorem{definition}{Definition}
\newtheorem{lemma}{Lemma}
\newtheorem{assumption}{Assumption}
\newtheorem{assumptionLL}{Assumption}
\renewcommand\theassumptionLL{LL\arabic{assumptionLL}}

\theoremstyle{definition}
\newtheorem{example}{Example}
\newtheorem{remark}{Remark}

\bibliographystyle{ecta}
 \setlength{\bibsep}{1pt}
\maketitle 

\begin{abstract}
	
	We propose new confidence sets (CSs) for the regression discontinuity parameter in fuzzy designs. Our CSs are based on   local linear regression, and are  bias-aware, in the sense that they  take possible  bias explicitly into account. Their construction shares 
	similarities with that of Anderson-Rubin CSs in exactly identified instrumental variable models, and thereby avoids issues with ``delta method'' approximations that underlie most commonly used  existing inference methods for fuzzy regression discontinuity analysis. Our CSs are asymptotically equivalent to existing procedures in canonical settings with strong identification and a continuous running variable. However, due to their particular construction they are also valid under a wide range of empirically relevant conditions in which existing methods can fail, such as setups with discrete running variables, donut designs, and weak 	identification.

\end{abstract}

\bigskip

\onehalfspacing

\section{Introduction}

Regression discontinuity designs can deliver credible identification of treatment effects from observational data in settings where the probability of receiving the treatment changes discontinuously with a running variable at some known threshold value. Such designs are called sharp (SRD) if the probability changes from zero to one, and fuzzy (FRD) otherwise. With both types of designs, methods based on local linear regression  are widely used  in empirical practice for estimation and inference \citep[e.g.][]{hahn2001identification}. 

The confidence intervals (CIs) typically reported in empirical FRD studies are obtained by applying techniques for handling smoothing bias, such as robust bias correction \citep{calonico2014robust} or  bias-aware critical values \citep{armstrong2018optimal}, to a delta method (DM)  approximation of the FRD estimator. Such DM CIs can be unreliable in practice, however, if  the running variable is not continuously distributed with full support around the cutoff, or the jump in treatment probabilities is ``small''. These limitations are important because empirical researchers often face running variables like   test scores or class sizes that take only a limited number of distinct values, ``donut designs''  that exclude  units close to the cutoff  to increase the credibility of causal estimates, or  weakly identified setups where treatment assignment only has a moderate impact on treatment probabilities.\footnote{To illustrate the scope of the issue, we surveyed the articles published between 2015 and 2021 in the ``Top 5'' economics journals. We found 20 papers that used a fuzzy regression discontinuity design as one of their main empirical specification; 9 of which had an ``irregular support'' (in the sense of having less than 100 support points within the bandwidth window on each side of the cutoff), and one was potentially affected by weak identification (in the sense that the reported first stage estimate differed by less than three standard errors from zero).}

In this paper, we propose  a new class of FRD confidence sets (CSs) that are not subject to these issues. The idea is to apply a bias-aware approach, which takes possible finite sample biases into account, to a particular local linear SRD estimator that is conceptually similar to an Anderson-Rubin statistic in a linear IV model \citep{staiger1997instrumental,feir2016weak}.
We show that the resulting CSs are ``honest'' in the sense of \citet{li1989honest}, meaning that they have correct asymptotic coverage uniformly over a class of conditional expectation functions of outcomes and treatments with bounded second derivatives, irrespective of the distribution of the running variable or the strength of identification. We also show that  our CSs are asymptotically equivalent
to  bias-aware DM CIs in settings with a continuous running variable  and strong identification.

Regression discontinuity methods that explicitly take possible bias into account have been shown to have favorable theoretical and practical properties, for instance, by \citet[]{armstrong2018optimal, armstrong2018simple}, \citet{kolesar2018discrete} and \citet{imbens2019optimized}. The CSs in this paper complement these methods by providing reliable FRD inference under potentially challenging circumstances without sacrificing efficiency in the canonical setup. Our approach is related to that of \citet{feir2016weak}, who also consider Anderson-Rubin-type statistics in FRD designs with potentially small jumps in treatment probabilities, but differs in that it allows for discrete (or otherwise irregularly supported) running variables, takes potential bias explicitly into account, and includes a method for choosing bandwidths in practice.

\section{Setup and Preliminaries}\label{sec:setup}

 Let $Y_i\in\IR$  be the outcome, $T_i\in\{0,1\}$  the actual 
treatment status, $Z_i\in\{0,1\}$  the assigned treatment, and $X_i\in\IR$  the running variable
of the $i$th unit in a random sample of size $n$ from a large population. Treatment is
assigned if the running variable falls above a known cutoff that we normalize  to zero, so that  $Z_i = \1{X_i \geq 0}$. The parameter of interest is $\poi  = \tau_Y/\tau_T,$
where for a generic  random variable $W_i$ we write $\mu_W(x)=\E(W_i|X_i=x)$ for is its conditional expectation
function given the running variable; $\mu_{W+} =\lim_{x\downarrow 0}\mu_W(x) $ and $\mu_{W-} =\lim_{x\uparrow 0}\mu_W(x)$ for the right and left limit at
the cutoff; and $\tau_W =\mu_{W+}-\mu_{W-}$ for the corresponding jump.\footnote{We write
$f_{+} =\lim_{x\downarrow 0}f(x) $ and $f_{-} =\lim_{x\uparrow 0}f(x)$ for generic functions $f$ throughout the paper.} 
In a potential outcomes framework with certain continuity and monotonicity conditions \citep[e.g.][]{hahn2001identification},  the parameter $\poi$ has a causal interpretation as the local average treatment effect 
among units at the cutoff whose treatment decision is affected
by the assignment rule. 

Our goal  is to construct powerful confidence sets    $\mathcal{C}^\alpha$ that  cover the parameter $\poi$ in large samples with at least probability $1-\alpha$, uniformly over $(\mu_Y,\mu_T)$ in some  function class $\mathcal{F}$ that embodies shape restrictions imposed by the analyst:\footnote{Note that  we leave the dependence of the probability measure $\IP$ and
	the parameter $\poi$ on $\mu_Y$ and $\mu_T$ implicit in our notation. Each function pair $(\mu_Y,\mu_T)$ corresponds to a single distribution of  $(Y,T,X,Z)=(\mu_Y(X)+\epsilon_M,\1{\mu_T(X)\geq\epsilon_T},X,Z)$,
	where $(\epsilon_M,\epsilon_T)$
	is some fixed  random vector. } 
\begin{align}
\liminf_{n\to\infty} \inf_{(\mu_Y, \mu_T)\in\mathcal{F}}\IP(\poi \in \mathcal{C}^\alpha) \geq 1-\alpha. \label{honest}
\end{align}
Following \citet{li1989honest}, we refer to such CSs  as \emph{honest with respect to $\mathcal{F}$}. This is a stronger
 requirement than correct pointwise asymptotic coverage:
\begin{align}
\liminf_{n\to\infty} \IP(\poi \in \mathcal{C}^\alpha) \geq 1-\alpha \textnormal{ for all } (\mu_Y, \mu_T)\in\mathcal{F}. \label{pointwise}
\end{align}
In particular, under~\eqref{honest} we can always find a sample size $n$ such that the coverage
probability of $\mathcal{C}^\alpha$ is not below $1-\alpha$ by more than an arbitrarily small
amount for every $(\mu_Y, \mu_T)\in\mathcal{F}$.  Under~\eqref{pointwise} there is no such guarantee, and even in very large samples the coverage probability
of $\mathcal{C}^\alpha$ could be poor for some $(\mu_Y, \mu_T)\in\mathcal{F}$. Since we do not know in advance which function pair is the correct
one, honesty as in~\eqref{honest} is necessary for good finite sample coverage
of $\mathcal{C}^\alpha$ across data generating processes.

As in \citet{imbens2019optimized} or \citet{armstrong2018simple}, we specify   $\mathcal{F}$   as a smoothness class.  Specifically, let
$\mathcal{F}_H(B)=\{f_1(x)\1{x\geq 0}- f_0(x)\1{x< 0}: \|f_w''\|_{\infty} \leq B, w=0,1\}$ be the Hölder-type class of real functions that are potentially discontinuous at zero, are twice
differentiable on either 
side of the threshold, and whose  second derivatives  are uniformly bounded by some constant $B>0$;  and  let
$\mathcal{F}_H^\delta(B)=\{f\in\mathcal{F}_H(B): |f_+ - f_-|>\delta\}$
 be a similar class of functions whose jump at zero is larger than some $\delta\geq 0$.  
We  then assume that there are constants  $B_Y$ and $B_T$, whose choice we discuss in Section~\ref{sec:choose_bounds}, such that
\begin{align}
(\mu_Y,\mu_T) \in \mathcal{F}_H(B_Y) \times \mathcal{F}_H^0(B_T) \equiv \mathcal{F}.\label{ass_hclass}
\end{align}
As $\mathcal{F}$ is a Cartesian product, this rules out cross-restrictions between $\mu_Y$ and~$\mu_T$. Note that we impose $\mu_T\in \mathcal{F}_H^0(B_T)$, and thus  $\tau_T\neq 0$,  only to ensure that $\poi=\tau_Y/\tau_T$ is well-defined. We explicitly allow  $\tau_T$ to be arbitrarily close to zero.

If the running variable is discrete, or more generally such that there are gaps in its support, condition~\eqref{ass_hclass} is understood to mean that there exists a  function pair  $(\mu_Y,\mu_T) \in \mathcal{F}$  such that $(\mu_Y(X_i),\mu_T(X_i))=(\E(Y_i|X_i),\E(T_i|X_i))$ with probability one \citep[cf.][]{kolesar2018discrete}. With this interpretation, the parameter $\theta$ is generally  partially identified:
\begin{multline*}
	\poi\in\Theta_I \equiv \left\{\frac{m_{Y+}-m_{Y-}}{m_{T+}-m_{T-}} : (m_Y,m_T)\in \mathcal{F}, \textnormal{ }
	(m_Y(X_i),m_T(X_i))=(\E(Y_i|X_i),\E(T_i|X_i))\right\},
\end{multline*}
where the  identified set $\Theta_I$ is   either (i) a closed interval $[a_1,a_2]$ with $a_1\leq a_2$; (ii) the union of
two disjoint half-lines, $(-\infty,a_1]\cup[a_2,\infty)$ with  $a_1<0<a_2$; (iii) the entire real line; or, as a
knife-edge case (iv) a half-line $[a_1,\infty)$ or $(-\infty,-a_1]$, with $a_1>0$.\footnote{This  holds because the range of $(m_{Y+}-m_{Y-},m_{T+}-m_{T-})$ over   $(m_Y,m_T)\in \mathcal{F}$  is a Cartesian product   of two intervals $I_Y \times I_T$. The four
	cases then obtain depending on which of these two intervals contain zero, possibly as
	a boundary value. } 
 The classical point identification result when the support of $X_i$ contains an open neighborhood around the cutoff  is then simply  a special case of (i) with   $\Theta_I$ a singleton. Our goal is to construct CSs for $\poi$ that have correct uniform asymptotic coverage under both  point and partially identification \citep[cf.][]{imbens2004confidence}, without applied researchers having to decide which of the two notions of identification more accurately applies to their specific setting.

\section{Bias-Aware Anderson-Rubin-Type Confidence Sets}\label{section_cs}

We argue in Section~\ref{subsection:frdinference} that conventional CIs,  based on local linear regressions \citep{fan1996local} and ``delta method'' (DM)  arguments, can potentially  break down in a number of practically relevant setups, including ones with discrete running variables or weak identification. Our proposed  approach is still based on local linear regression, but avoids these issues through  a construction similar to that of \cite{anderson1949estimation} for inference in exactly identified linear IV models. It also takes possible bias from local linear smoothing explicitly into account. We hence refer to our CSs  as  \emph{bias-aware AR CSs}. 

To describe the approach, we write $\widehat\tau_W(h)$ for the local linear estimator  of the jump $\tau_W$ in the conditional expectation of some generic random variable $W_i$ given the running variable $X_i$ at the cutoff:
\begin{align}
	\widehat\tau_W(h) &=  e_1^\top\argmin_{\beta\in\IR^4} \sum_{i=1}^n K(X_i/h) (W_i -  \beta^\top (Z_i, X_i, Z_i X_i,1))^2  =\sum_{i=1}^n  w_{i}(h) W_i.\label{sharpRDDest}
\end{align}
Here  $K(\cdot)$  is a  kernel function, $h>0$ is a bandwidth,
 $e_1 = (1,0,0,0)^\top$ is  the first unit vector, and the  $w_i(h)$ are weights, given explicitly in Appendix~\ref{section:proofsofmainresults}, that only depend on the data through the realizations $\mathcal{X}_n = (X_1,\ldots,X_n)^\top$ of the running variable. We refer to estimators of the form in~\eqref{sharpRDDest} as \emph{SRD-type estimators
	of $\tau_W$} in the following.

The natural point estimator  of $\poi$ is $\widehat\poi(h)  = \widehat\tau_Y(h)/\widehat\tau_T(h)$, but we will base inference on different statistics. Define the auxiliary parameters
$\tau_{M}(c)=\tau_{Y}- c\tau_{T}$ for  $c\in\IR$, and note that $\tau_{M}(c) = \mu_{M+}(c)-\mu_{M-}(c)$, with $\mu_{M}(x,c) =\E(M_i(c)|X_i=x)$ and  $M_i(c) = Y_i - cT_i$. We then consider the SRD-type estimator $\widehat\tau_{M}(h,c)=\sum_{i=1}^n  w_{i}(h) M_i(c)$, and 
exploit the properties of the regression weights $w_i(h)$ to write its conditional bias $b_{M}(h,c)=\E(\widehat\tau_{M}(h,c)|\mathcal{X}_n)-\tau_{M}(c)$ and conditional variance $s^2_{M}(h,c) =\V(\widehat\tau_{M}(h,c)|\mathcal{X}_n)$ 
 given  $\mathcal{X}_n$ as
\begin{align*}
b_{M}(h,c) = \sum_{i=1}^n w_i(h)\mu_M(X_i,c) - ( \mu_{M+}(c)-\mu_{M-}(c)), \quad
s^2_{M}(h,c) = \sum_{i=1}^n  w_i(h)^2 \sigma_{M,i}^2(c),
\end{align*}
respectively, with  $\sigma_{M,i}^2(c) = \V(M_i(c)|X_i)$  the conditional variance of $M_i(c)$ given $X_i$.\footnote{To keep the notation simple, the estimator  $\widehat\tau_{M}(h,c)=\widehat\tau_{Y}(h)- c\widehat\tau_{T}(h)$ uses the same bandwidth
	on each side of the cutoff, and also the same bandwidth for estimating $\tau_Y$ and $\tau_T$. It is straightforward to accommodate more general bandwidth
	choices; see Online Appendix~\ref{sec:appendixC} for details.} 

The bias depends on  $(\mu_Y,\mu_T)$ through
the transformation $\mu_M = \mu_Y - c  \mu_T$ only, and we have that $\mu_M \in \mathcal{F}_H(B_Y+|c| B_T)$ by~\eqref{ass_hclass}.
As in \citet{armstrong2018simple},  for any value of the bandwidth $h$ we can thus explicitly  bound $b_{M}(h,c)$ in absolute value
 over  $\mathcal{F}$:
\begin{align*}
\sup_{(\mu_Y,\mu_T)\in\mathcal{F}}| b_{M}(h,c)| \leq  \overline{b}_{M}(h,c)\equiv  - \frac{B_Y + |c|  B_T}{2}  \sum_{i=1}^n w_{i}(h) X_i^2  \sgn(X_i).
\end{align*}
The supremum is achieved by the ``worst case'' pair of piecewise quadratic conditional expectation function with second derivatives equal to $(B_Y \textnormal{sign}(x), - B_T \textnormal{sign}(x))$ over  $x\in[-h,h]$.\footnote{
	Note that this bound may not
	be sharp if no such pair of piecewise quadratic functions 
	is  a feasible candidate for $(\mu_Y,\mu_T)$. For example, there is no function $\mu_T$ with $\mu_T''(x)= B_T\textnormal{sign}(x)$ and $\mu_T(x)\in [0,1]$ for all $x\in[-h,h]$ if $h> (2 / B_T)^{1/2}$. 
	Still, the bias bound is valid in such cases.}

For every $c\in\IR$  we can then construct an infeasible  bias-aware CI for  
$\tau_{M}(c)$ as 
$$C_M^{\alpha}(h,c) =\left[\widehat\tau_{M}(h,c) \pm\textnormal{cv}_{1-\alpha}(  r_{M}(h,c))   s_{M}(h,c)\right],$$ 
where $r_{M}(h,c) = \overline{b}_{M}(h,c)/s_{M}(h,c)$ is
 the ``worst case'' bias to 
 standard deviation ratio, and  $\textnormal{cv}_{1-\alpha}(r)$ is  the $(1-\alpha)$-quantile of  ``folded''  normal distribution
 $|N(r,1)|$.
 \citet{armstrong2018optimal,armstrong2018simple} show that such CIs are honest with respect to $\mathcal{F}_H(B_W)$ irrespective of the distribution of the running variable, have correct asymptotic coverage $1-\alpha$ at the ``worst case'' conditional expectations, are valid for wide ranges of bandwidths, and are highly efficient for SRD inference if the  running variable is continuous.

    The bandwidth that minimizes this CI's asymptotic length is
$$\hopt(c) = \argmin_{h}  \textnormal{cv}_{1-\alpha}( r_{M}(h,c))  s_{M}(h,c).$$
We assume that this optimal  bandwidth is unique; 
and the minimization in its definition is understood to be carried out over the set of bandwidths for which the involved objects are well-defined.
An efficient but infeasible bias-aware AR CS for $\poi$ is then given by  the set of all $c\in\IR$ for which the
auxiliary CI $C_M^{\alpha}(\hopt(c),c)$ contains the value zero:
 \begin{align}
\mathcal{C}_{*}^{\alpha}
=\left\{c : |\widehat\tau_{M}(\hopt(c),c)| \leq \textnormal{cv}_{1-\alpha}( r_{M}(\hopt(c),c)) 
s_{M}(\hopt(c),c) \right \}.\label{our_ci_infeas}
\end{align}
Our proposed bias-aware AR CSs are  feasible versions of~\eqref{our_ci_infeas} based on a standard error $\widehat s_{M}(h,c)= \sum_{i=1}^nw_i(h)^2\widehat\sigma_{M,i}^2(c)$ and  some estimate
 $\hhopt(c)$ of the optimal bandwidth:
 \begin{align}
\mathcal{C}_{\textnormal{ar}}^{\alpha}
=\left\{c : |\widehat\tau_{M}(\hhopt(c),c)| \leq \textnormal{cv}_{1-\alpha}(\widehat r_{M}(\hhopt(c),c)) 
\widehat{s}_{M}(\hhopt(c),c)) \right \},\label{our_ci}
\end{align}
with $\widehat r_{M}(h,c) = \overline{b}_{M}(h,c)/\widehat s_{M}(h,c)$. Both the standard error and bandwidth estimator can be implemented in different ways, and our theoretical analysis below therefore
only imposes some weak ``high level'' conditions.
We  propose a specific   standard error  based on  
nearest-neighbor linear regression estimates $\widehat\sigma_{M,i}^2(c)$ of  $\sigma_{M,i}^2(c)$ in Section~\ref{subsectionstderrors}; and  a feasible bandwidth  that combines a plug-in construction
with a   safeguard against certain small sample distortions in
Section~\ref{sec:bandwidth_choice}.

\section{Theoretical Properties}\label{subsection:theoreticalpropoerties}

\subsection{Coverage}\label{subsection:ass} Our main theoretical result is that $ \mathcal{C}_{\textnormal{ar}}^{\alpha}$ 
is an honest CS for $\poi$ with respect to $\mathcal{F}$, in the sense of~\eqref{honest}, under the following rather weak conditions.

\begin{assumption}\label{reg1} (i) The data $\{(Y_i,T_i,X_i),i=1,\ldots,n\}$ are an  i.i.d. sample; (ii) $\E( (Y_i - \E(Y_i|X_i))^q| X_i=x)$ exists and 
	is bounded uniformly over $x\in\supp(X_i)$ and $(\mu_Y,\mu_T)\in\mathcal{F}$
	for some $q>2$;
	(iii) $\V (Y_i|X_i=x)$ is bounded away from zero uniformly over $x\in\supp(X_i)$ and $(\mu_Y,\mu_T)\in\mathcal{F}$; and  $\Cov(Y_i, T_i|X_i=x)^2/(\V (Y_i|X_i=x) \V (T_i|X_i=x))$ is
	bounded away from one uniformly over $x\in\supp(X_i)\cup\{x: \V (T_i|X_i=x)>0 \}$ and $(\mu_Y,\mu_T)\in\mathcal{F}$; (iv) the kernel function $K$ is a continuous, unimodal, symmetric density
	function that is equal to zero outside some compact set, say $[-1,1]$.
\end{assumption}
Assumption~\ref{reg1} collects mostly standard conditions from the literature on
local linear regression. Part~(i) could be weakened to allow for certain forms of dependent sampling,
such as cluster sampling.  Parts~(ii) and~(iii) ensure that $\V(M_i(c)|X_i=x)$ is bounded away
from zero for all $c\in\IR$ and allow for the special case of a SRD design. Part~(iv) is satisfied by most kernel
functions commonly used in applied RD analysis, such as the triangular or the Epanechnikov kernels.

\begin{assumption}\label{reg2} The following holds uniformly over $(\mu_Y,\mu_T)\in\mathcal{F}$: (i)   $\hhopt(c)=\hopt(c)(1 +o_P(1))$; and (ii)
	 $\widehat{s}_{M}( \hhopt(c),c) = s_{M}( h_{M}(c),c)(1+o_P(1))$.
\end{assumption}

Part~(i) of Assumption~\ref{reg2} states that the empirical bandwidth is consistent
for the infeasible optimal one, and part (ii) states that the empirical standard error  is  consistent for the true standard deviation at the infeasible optimal bandwidth. We discuss specific implementations in Sections~\ref{subsectionstderrors} and~\ref{sec:bandwidth_choice}.

\begin{assumptionLL}
	\label{assumptiondiscrete}	The support of  the running variable $X_i$ is finite and symmetric, in the sense that it is of the form  $\{\pm x_1, \ldots, \pm x_k\}$, for positive constants $(x_1,\ldots,x_k$) over some open neighborhood 	of the cutoff. 
\end{assumptionLL}

\begin{assumptionLL}\label{assumptioncontinuous}
	 The running variable  $X_i$ is continuously distributed with continuous density $f_X$ that is bounded and bounded away from zero over an open neighborhood of the cutoff. 
\end{assumptionLL}

Assumptions~\ref{assumptiondiscrete}--\ref{assumptioncontinuous} describe RD setups with
discrete and continuously distributed running variables, respectively. These settings are meant to be exemplary and are considered because they allow explicit characterization of the bandwidth $h_M(c)$.
 Note that the symmetry of the support in Assumption~\ref{assumptiondiscrete} is  for notational convenience only. Discrete running variables with asymmetric support can easily be accommodated by using a different bandwidth on each side of the cutoff, as  described in Online Appendix~\ref{sec:appendixC}. In Appendix~\ref{app:alternative} we also consider an alternative asymptotic framework for the discrete case.

 In
Lemma~\ref{theorem_assumptions} in the Appendix, we show that our assumptions have two main implications:
(i) using an estimate of the optimal bandwidth instead of its
population version has a small impact, in some appropriate sense, on the quantities involved in the construction of our CS; (ii)  the magnitude of each  weight $w_i(h_{M}(c))$ is small relative to the others' in large samples, in the sense that 
$w_\textnormal{ratio}(h) \equiv \max_{j=1,\ldots,n} w_j(h)^2/\sum_{i=1}^n w_i(h)^2=o_P(1),$ so that a CLT
applies to an appropriately standardized version of the estimator of $\tau_{M}(c)$.
This yields the following formal result.

\begin{theorem}\label{th1}Suppose that Assumptions~\ref{reg1}--\ref{reg2} and either \ref{assumptiondiscrete} or \ref{assumptioncontinuous} hold. Then $\mathcal{C}_{\textnormal{ar}}^{\alpha}$
	is honest with respect to $\mathcal{F}$ in the sense of 	\eqref{honest}.
\end{theorem}

\subsection{Shape}

Because our CS $\mathcal{C}_{\textnormal{ar}}^{\alpha}$ is defined through an inversion argument,
it is interesting to study its shape. A simple sufficient condition for $\mathcal{C}_{\textnormal{ar}}^{\alpha}$  to be non-empty
is that the bandwidth $\hhopt(c)$ is continuous in $c$, but beyond that it is difficult to make general
statements. To see why, recall that $c\in \mathcal{C}_{\textnormal{ar}}^{\alpha}$ if and only if
\begin{align*}
|\widehat\tau_{M}(\widehat h_{M}(c),c)|\leq \textnormal{cv}_{1-\alpha}(\widehat r_{M}(\widehat h_{M}(c),c))
\widehat{s}_{M}(\widehat h_{M}(c),c).
\end{align*}
The above quantities depend on $c$ directly, but also indirectly through $\widehat h_{M}(c)$. While the former dependence is rather simple in structure, the latter introduces complicated nonlinearities that make it impossible to give a simple analytical
result regarding the shape of our CS.
Such a result
is possible, however, for a version that uses a fixed bandwidth.
\begin{theorem}\label{theorem_formofCS}
	Let $ \mathcal{C}_{\textnormal{ar}}^{\alpha}(h)$ be a
	version of $\mathcal{C}_{\textnormal{ar}}^{\alpha}$ that uses a  bandwidth 	$h$  that does not depend on $c$. Then either  $\; \mathcal{C}_{\textnormal{ar}}^{\alpha}(h)=[a_1,a_2]$, or 
	$\; \mathcal{C}_{\textnormal{ar}}^{\alpha}(h)=(-\infty,a_1]\cup[a_2,\infty)$, or $\; \mathcal{C}_{\textnormal{ar}}^{\alpha}(h)=(-\infty,\infty)$, or $\; \mathcal{C}_{\textnormal{ar}}^{\alpha}(h)=[a_1,\infty)$ or $\; \mathcal{C}_{\textnormal{ar}}^{\alpha}(h)=(-\infty,a_1]$, for some constants $a_1<a_2$.
\end{theorem}

The result mirrors the discussion at the end of Section~\ref{sec:setup}. It suggests that our actual CS   should  take one of these
general shapes as long as  $\widehat h_{M}(c)$ does not vary  ``too  much'' with $c$. 
	We found this to be the case in 
	every  simulation run and every empirical analysis that we conducted in the context
	of this paper. The last two cases in Theorem~\ref{theorem_formofCS}, in which $\mathcal{C}_{\textnormal{ar}}^{\alpha}(h)$ is a
half-line, are also ``knife-edge'' cases: they only occur if one of the boundaries of a bias-aware CI for $\tau_T$ is exactly equal to zero, and are thus largely irrelevant for empirical practice.

\section{Comparison with Delta Method Inference}\label{subsection:frdinference}

\subsection{Method and Limitations} The CIs  commonly reported in empirical FRD studies are based on a linearization or ``delta method'' (DM)  argument. It starts by noting that, after centering, the FRD point estimator  $\widehat\poi(h)  = \widehat\tau_Y(h)/\widehat\tau_T(h)$ can be written as 
the sum of an SRD-type estimator $\widehat\tau_U(h)$ with unobserved dependent variable $U_i$, and a remainder $\widehat\rho(h)$:
\begin{align*}
&\hpoi(h)-\poi=\widehat\tau_U(h) + \widehat\rho(h), \quad \widehat\tau_U(h) = \sum_{i=1}^nw_i(h)U_i, \quad  U_i = \frac{Y_i-\tau_Y}{\tau_T} - \frac{\tau_Y (T_i-\tau_T)}{\tau_T^2},\\ &\qquad\widehat\rho(h)=\frac{\widehat\tau_Y(h)(\widehat\tau_T(h)-\tau_T)^2}{2\widehat\tau_T^*(h)^3}-\frac{(\widehat\tau_Y(h)-\tau_Y)(\widehat\tau_T(h)-\tau_T)}{ \tau_T^2},
\end{align*}
with  $\widehat\tau_T^*(h)$  an intermediate value between $\tau_T$ and $\widehat\tau_T(h)$. One then imposes  conditions under which 
$\widehat\rho(h)$ is asymptotically negligible relative to $\widehat\tau_U(h)$, and forms a CI for $\poi$ 
by applying some method for SRD inference to $\widehat\tau_U(h)$, which differ mainly in how they handle potential bias.
  Such DM CIs are proposed, for example, by \citet{calonico2014robust} and \citet{armstrong2018simple} in combination with robust bias correction and a  bias-aware approach, respectively.\footnote{In empirical papers, FRD estimates are sometimes
	obtained through the two-stage least squares regression
	$Y_i = \theta T_i + \beta_+ X_iZ_i + \beta_-X_i(1-Z_i) + \varepsilon_{i}$    with $Z_i$ as an instrument for $T_i$,
	using only data in some window around the cutoff. This is numerically equivalent to a ratio of local
	linear regressions with a uniform kernel, and the resulting CI is thus of the DM type  \citep{hahn2001identification,imbens2008regression}.} As $U_i$ is unobserved, any
such method must also be made feasible by using an estimate $\widehat U_i$ in which $\tau_Y$ and $\tau_T$ are replaced by suitable preliminary estimators.

Obvious downsides of such constructions include that they only control the bias of a first-order approximation of $\hpoi(h)$, and
not the bias of $\hpoi(h)$ itself; and that replacing $U_i$ with an estimate $\widehat U_i$  introduces additional uncertainty,  which is asymptotically second-order and hence generally unaccounted for in practice. In FRD designs such CIs are therefore generally subject to additional finite-sample distortions, relative to SRD designs. 

A more principal  issue with DM CIs is that a central condition for their validity, namely  that $\widehat\rho(h)$ is asymptotically
negligible relative to $\widehat\tau_U(h)$,   is not innocuous. In particular, 
this condition is not compatible with
a discrete running variable, or more generally one with support gaps around the cutoff.
This is because  $\tau_T$  and $\tau_Y$ are generally only partially identified in this case, and hence cannot be consistently estimated; see Section~\ref{sec:setup}. The term  $\widehat\rho(h)$  then generally has a non-zero probability limit, and 
cannot be ignored for the purpose of inference on $\poi$. This issue occurs irrespective of the method chosen to control the bias of $\widehat\tau_U(h)$, including bias-aware
inference. Because  running variables with discrete or irregular support are ubiquitous in practice, this is an important limitation.

Another issue with DM CIs is that the conditions for their uniform validity   rule out weakly identified settings with $\tau_T$ close to zero. This problem occurs even if the running variable is continuously distributed, and with any method chosen to control the bias of $\widehat\tau_U(h)$, including bias-aware
inference. This is because for any DM CI to be honest with respect
to $\mathcal{F}$, the term $\widehat\rho(h)$ must be of smaller order than $\widehat\tau_U(h)$ not only at the ``true''
function pair $(\mu_Y,\mu_T)$, but uniformly over all $(\mu_Y,\mu_T)\in\mathcal{F}$. But since
$\tau_T$ can be arbitrarily close to zero over $(\mu_Y,\mu_T)\in\mathcal{F}$,  we  have that
$\sup_{(\mu_Y,\mu_T)\in\mathcal{F}}| \widehat\rho(h)| = \infty$, which means that DM CIs can be unreliable in such settings.\footnote{\citet{feir2016weak} also point out coverage issues of
	DM CIs under weak identification, but use  different types of arguments. Specifically, they show 
	that DM CIs based on infeasible ``undersmoothing'' bandwidths do not have correct asymptotic coverage under
	pointwise (with respect to the involved conditional expectation functions) asymptotics if $\tau_T$ tends
	to zero at an appropriate rate related to that of the bandwidth. They also show that undersmoothing AR CSs can have correct poinwise asymptotic coverage in this case.}

\subsection{An Equivalence Result}
\label{sec:biasawaredeltamethod}

\citet{armstrong2018simple} study bias-aware DM CIs  under conditions for which such DM 
CIs are asymptotically valid. These include Assumption~\ref{assumptioncontinuous},
which implies that is $X_i$ continuously distributed, and  that 
$(\mu_Y,\mu_T) \in \mathcal{F}_H(B_Y) \times \mathcal{F}_H^\delta(B_T) \equiv \mathcal{F}^\delta$ for some $\delta>0$, which means that
$\tau_T$ is
well-separated from zero. They  show that bias-aware DM CIs are honest
with respect to $\mathcal{F}^\delta$ in this case, and also near-optimal, in the sense that no
other method can substantially improve upon their length in large samples. 
The next theorem shows that our bias-aware AR CSs are as efficient as their DM counterparts in such settings for which DM CIs are specifically designed.

To avoid introducing additional high-level assumptions about the implementation details  of bias-aware DM CIs  we consider an infeasible  version $\mathcal{C}_{\Delta}^{\alpha}$, formally defined in \eqref{c_delta}, and compare
it to its infeasible counterpart $\mathcal{C}_{*}^{\alpha}$ in our setup. Equal efficiency
is established in the sense
that both  CSs have the same local asymptotic coverage for a drifting parameter within a $O(n^{-2/5})$ neighborhood of $\poi$. Such neighborhoods are appropriate to consider as the length of $\mathcal{C}_{\Delta}^{\alpha}$ is $O_P(n^{-2/5})$ uniformly over $\mathcal{F}^\delta$.

\begin{theorem}\label{theorem_delta_ar}
	Suppose that Assumptions~\ref{reg1}--\ref{reg2} and~\ref{assumptioncontinuous} hold,
	and put $\poi^{(n)}=\poi + \kappa   n^{-2/5}$ for some
	constant $\kappa$. Then
	$$\limsup_{n\to\infty} \sup_{(\mu_Y,\mu_T)\in\mathcal{F}^\delta}\left|	\IP\left(\poi^{(n)}\in  \mathcal{C}_{*}^{\alpha}\right) -	\IP\left(\poi^{(n)}\in  \mathcal{C}_{\Delta}^{\alpha}\right)\right|=0.$$
\end{theorem}

This result parallels the well-known finding that there is no loss of efficiency when using the AR approach  in exactly identified IV models relative to one based on a conventional
$t$-test \citep[e.g.][]{andrews2019weak}. It is not an obvious corollary, however, as there are, for example, no analogues to the bandwidth and the smoothing bias in such IV models.

\section{Implementation Details and Extensions}\label{sectionextensions}

\subsection{Standard Errors} \label{subsectionstderrors}

Natural  standard errors for $\widehat\tau_M(h,c)$ are
of the form
$\widehat s_{M}(h,c) = (\sum_{i=1}^n  w_i(h)^2 \widehat\sigma_{M,i}^2(c))^{1/2}$,
 with  $\widehat\sigma_{M,i}^2(c)$ some estimate of $\sigma_{M,i}^2(c)$. 
Setting $\widehat\sigma_{M,i}^2(c)$ to the squared difference
between the outcome  of unit $i$ and the  average outcome among its nearest neighbors in terms of the running variable
 \citep{abadie2006large,abadie2014inference} is commonly
recommended in the RD literature \citep[e.g.][]{calonico2014robust}. However, this nearest-neighbor standard error is actually not  uniformly consistent over $\mathcal{F}$ because the leading bias of $\widehat\sigma_{M,i}^2(c)$ is proportional to the first derivative of $\mu_M(\cdot,c)$ at $X_i$, which
is  unbounded over $\mathcal{F}$. 
We therefore propose  a novel  procedure that replaces the local
sample average with a local best linear predictor. This modification makes the bias of $\widehat\sigma_{M,i}^2(c)$ proportional to the  second derivative of $\mu_M(\cdot,c)$ at $X_i$, which is bounded in absolute value over $\mathcal{F}$ by $B_Y+|c| B_T$. 

We propose a version that explicitly allows for ties among the realizations of the running variable. For $R$ a small integer,  denote the rank of $|X_j -X_i|$ among the elements of the set $\{|X_{s} -X_i|: s\in\{1,\ldots,n\}\setminus\{i\}, X_{s} X_i > 0\}$ by $r(j,i)$,
let $\mathcal{R}_i$ be the set of indices such that $r(j,i) \leq Q_i$, where $Q_i$ is the smallest
integer such that $\mathcal{R}_i$ contains at least $R$ elements,
and let  $R_i$
be the resulting cardinality of $\mathcal{R}_i$.\footnote{Note that if every realization of $X_i$ is unique, then $R=Q_i=R_i$, and $\mathcal{R}_i$
	is the set of unit $i$'s $R$ nearest neighbors' indices; but with ties in the data 
	$R_i$ could be greater than $R$.  We use $R=5$ in our simulations and the empirical application.  }
The estimator $\widehat{\sigma}^2_{M,i}(c)$ is then defined as the scaled squared difference between $M_i(c)$ and its best linear
predictor given its $R_i$ nearest neighbors:
\begin{align*}
&\widehat{\sigma}^2_{M,i}(c) = 
\frac{1}{1+ H_i}  \left( M_i(c)  - \widehat M_i(c)   \right)^2, \textnormal{ with}\\
&\widehat M_i(c)=\widetilde{X}_i\left(\sum_{j \in \mathcal{R}_i} \widetilde{X}_j^\top \widetilde{X}_j\right)^{-1} \sum_{j \in \mathcal{R}_i} \widetilde{X}_j^\top M_j(c), \quad H_i =  \widetilde{X}_i \left(\sum_{j \in \mathcal{R}_i} \widetilde{X}_j^\top \widetilde{X}_j \right)^{-1} \widetilde{X_i}^\top.
\end{align*}
Here $\widetilde{X}_i = (1,X_i)^\top$ if the running variable takes at least two distinct
values among the $R_i$ nearest neighbors of unit $i$, and $\widetilde{X}_i = 1$ otherwise.
 The   scaling term $H_i$ ensures that $\widehat{\sigma}^2_{M,i}(c)$ is approximately unbiased  in large samples. 
The next result, which we prove in Online Appendix~\ref{proof:consistentcyofse}, shows that our
new standard error is indeed uniformly consistent under general conditions.  We  recommend its use not just for our CS, but more generally for bias-aware inference methods that work with bounds on second derivatives.

\begin{theorem}\label{theorem_se} Suppose that Assumption~\ref{reg1}, Assumption~\ref{reg2}(i), and either Assumption~\ref{assumptiondiscrete}  or Assumption~\ref{assumptioncontinuous} are satisfied; that $\V (Y_i|X_i=x)$, $\V (T_i|X_i=x)$ and $\Cov(Y_i, T_i|X_i=x)$ are  Lipschitz continuous on each side of the cutoff uniformly over $x\in\supp(X_i)$ and $(\mu_Y,\mu_T)\in\mathcal{F}$; and that $\E( (Y_i - \E(Y_i|X_i))^4| X_i=x)$  is uniformly bounded over $x\in\IR$ and $(\mu_Y,\mu_T)\in\mathcal{F}$. Then 	
	Assumption~\ref{reg2}(ii) holds for the standard error described in this subsection.
\end{theorem}

\subsection{Bandwidth Choice}\label{sec:bandwidth_choice}
An obvious candidate for a feasible bandwidth is the empirical analogue of $\hopt(c)$, which  minimizes the length of the auxiliary CI in Section~4:
$$\hhopt^* (c) = \argmin_{h}  \textnormal{cv}_{1-\alpha}(\widehat r_{M}(h,c))  \widehat s_{M}(h,c).$$
While this choice is generally attractive, it could lead to
coverage distortions if $B_Y + |c| B_T$ is very large relative to sampling uncertainty.
To see why, recall from the discussion at the end of Section~\ref{subsection:ass} that  $\widehat\tau_{M}(h,c) =\sum_{i=1}^n  w_{i}(h) M_i(c)$ is asymptotically normal if  $w_{\textnormal{ratio}}(h)\equiv \max_{j=1,\ldots,n} w_j(h)^2/\sum_{i=1}^n w_i(h)^2=o_P(1)$. 
Normality should thus be a ``good'' finite-sample approximation   if $w_{\textnormal{ratio}}(h)$
is ``close'' to zero (this reasoning also follows from a Berry-Esseen-type result).
However,
if  $B_Y + |c| B_T$ (and thus the worst-case bias) is large, then  $\widehat h_{M}^*(c)$ is typically small. The weights $w_i(\widehat h_{M}^*(c))$ then concentrate on  few observations close to the cutoff, $w_{\textnormal{ratio}}(\widehat h_{M}^*(c))$ is
 large,   and CLT approximations can  be inaccurate as $\widehat\tau_{M}(\widehat h_{M}^*(c),c)$ then effectively
 behaves like a sample average of a small number of observations.

To address this issue, we propose imposing a lower bound on the bandwidth, chosen such 
 that the  value 
of $w_{\textnormal{ratio}}(h)$ remains below some reasonable threshold constant $\eta>0$, which we set to $\eta=.075$ in our simulations and empirical application.\footnote{To motivate this choice,
	suppose  that $\mathcal{X}_n = \{\pm .02, \pm .04,\ldots,\pm 1\}$, that
	$K(t)=(1-|t|)\1{|t|<1}$ is the triangular kernel, and that $h=1$. Then $\widehat\tau_M(h,c)$ is a weighted least squares 
	estimator that gives positive weight to 50 observations on each side of cutoff, and $w_{\textnormal{ratio}}(h)\approx .075$. }
 $$\widehat h_{M}(c) = \max\left\{\widehat h_{M}^*(c), h_{\min}(\eta)\right\}, \quad h_{\min}(\eta) = \min\left\{h: w_{\textnormal{ratio}}(h) < \eta\right\}.$$

Under standard conditions like Assumption~\ref{assumptiondiscrete}  or~\ref{assumptioncontinuous} the lower bound on the bandwidth clearly never binds asymptotically, but imposing it can improve the finite-sample coverage of our CSs: as $\hhopt(c) \geq \hhopt^*(c)$, our construction trades off a possible increase in finite-sample bias against normality being a better finite-sample approximation. This improves coverage
  because our CSs explicitly account for the exact
 bias, but cannot capture deviations from normality.
 This idea can also be used for SRD inference, and more generally in all settings where  finite-sample accuracy
	of inference faces a similar ``bias vs.\ normality'' trade-off.
	For example, \citet{armstrong2018finite} use our approach
	in the context of inference on average treatment effects
	under unconfoundedness with limited overlap.

\subsection{Choosing Smoothness Bounds}\label{sec:choose_bounds}

In order to compute $\mathcal{C}_{\textnormal{ar}}^{\alpha}$, researchers  needs to choose the smoothness bounds $B_Y$ and $B_T$. Such bounds cannot be estimated consistently  without imposing strong additional assumptions; and without choosing such bounds it is generally not possible to conduct inference on $\poi$ that is both valid and informative, even in large samples \citep{low1997nonparametric,armstrong2018optimal,bertanha2016impossible}. Methods that seem to such choices still require restrictions on smoothness implicitly to attain approximately correct CI coverage in practice \citep{armstrong2018simple}.\footnote{For example, an undersmoothing SRD CI can only be expected to have approximately correct coverage if the bias of the local linear estimator is ``small'' relative to its standard error.
This can only be expected if the underlying function is ``close'' to linear, which is equivalent to its maximum second derivative being	``close'' to zero. A researcher that considers an undersmoothing SRD CI to be reliable has thus implicitly imposed a smoothness bound. An analogous argument applies to robust bias correction \citep{kamat2018nonparametric}. } Explicitly specifying $B_Y$ and $B_T$ makes it transparent on which assumptions the inferential statements are based.

Roughly speaking, ``small'' values of $B_Y$ and $B_T$ amount to the assumption that the respective functions  are   ``close'' to linear on either side of the cutoff, whereas  larger values allow the functions to be increasingly ``curved''. The choice should be guided by subject knowledge but is arguably difficult in empirical practice, where there will be no single objectively  correct  value. In line with the previous literature, we hence recommend 
considering a range of plausible values as a form of sensitivity analysis.
  We also recommend estimating lowers bound $\widehat{B}_{Y,\textnormal{low}}$ and $\widehat{B}_{T,\textnormal{low}}$, and to compute  
one-sided CIs for $B_Y$ and $B_T$, respectively, via the methods proposed in \citet{armstrong2018optimal} and \citet{kolesar2018discrete}, to guard against overly optimistic choices.

Two heuristic ``rules of thumb'' (ROT) for determining plausible values in practice have also been considered in the literature. Both are based on fitting global polynomial specifications $\widetilde \mu_{Y,k}$ and $\widetilde \mu_{T,k}$ of order $k$ on either side of the cutoff by conventional least squares.
 \citet{armstrong2018simple} use fourth-order polynomials, and propose the   ROT   $\widehat B_{Y,\textnormal{ROT1}}= \sup_{x\in\mathcal{X}}|\widetilde \mu_{Y,4}''(x)|$ and $\widehat B_{T,\textnormal{ROT1}}= \sup_{x\in\mathcal{X}}|\widetilde \mu_{T,4}''(x)|$, where $\mathcal{X}$ denotes the support of the running variable.  \citet{imbens2019optimized} consider a   ROT
in which the maximal curvature implied by a quadratic fit is multiplied by some moderate factor, say 2, yielding  $\widehat B_{Y,\textnormal{\textnormal{ROT2}}}=2\sup_{x\in\mathcal{X}}|\widetilde \mu_{Y,2}''(x)|$ and $\widehat B_{T,\textnormal{\textnormal{ROT2}}}=2\sup_{x\in\mathcal{X}}|\widetilde \mu_{T,2}''(x)|$.

 Such rules of thumb
can provide  useful first guidance, but should  be complemented
with other approaches in a sensitivity analysis. We strongly recommend to always
check the fit of the respective polynomial specification, and to dismiss the ROT
value if the fit is obviously poor.
In Online Appendix~\ref{sec:appendixB}, we argue that in ``roughly quadratic'' settings  the fourth-order polynomial specification that underlies  ROT1 tends to produce quite erratic over-fits of the data that can lead to vast over-estimates of the true smoothness bounds, and corresponding CSs with poor statistical power.  ROT2, on the other hand, tends
to produce more reasonable values in many such setups.  This pattern is also visible in our simulations.

\subsection{Regression Kink Designs}

In Online Appendix~\ref{section:rkd}, we present an extension of our approach to  CSs for the ratio $\theta^{(v)}\equiv\tau_Y^{(v)} /\tau_T^{(v)} \equiv (\mu_{Y+}^{(v)}- \mu_{Y-}^{(v)})/(\mu_{T+}^{(v)}- \mu_{T-}^{(v)})$ of the jumps in the $v$th-order derivatives of two generic conditional expectation functions $(\mu_Y,\mu_T)$ at some threshold value. This setup covers the important Fuzzy Regression Kink Design \citep{card2015inference}, where the
parameter of interest is the ratio $\theta^{(1)}$ of two jumps in   first derivatives.

We again  form bias-aware CIs for an auxiliary parameter $\tau^{(v)}_M(c) = \tau^{(v)}_M - c\tau^{(v)}_T$, now  based on  $p$th-order local polynomial regression (where $p\geq v$ and typically $p=v+1$), and  collect all values of $c$ for which such CIs contain zero to create an AR CS for $\theta^{(v)}$. The construction is largely analogous to that described in  Section~\ref{section_cs}, with the main difference concerning the bias bound. Specifically, we derive the apparently  
novel result   that if $\mu_Y$ and $\mu_T$ are both $(p+1)$ times continuously differentiable on either side of the threshold, with  derivatives of order $(p+1)$ uniformly bounded by $B_Y$ and $B_T$, respectively, and $\widehat\tau_{M,p}^{(v)}(h,c)=\sum_{i=1}^n w_{vp,i}(h)M_i(c)$ is    the local $p$th order polynomial  estimator of $\tau^{(v)}_M(c)$ with bandwidth $h$, the conditional bias $\E(\widehat\tau_{M,p}^{(v)}(h,c)|\mathcal{X}_n)-\tau_{M,p}^{(v)}(c)$ is  absolutely bounded by
\begin{align*}
 \overline{b}_{M,vp}(h,c)\equiv   (-1)^{p-v} \frac{B_Y + |c|  B_T}{(p+1)!} \sum_{i=1}^n w_{vp,i}(h) X_i^{p+1}  \sgn(X_i)^{v+1},
\end{align*}
uniformly over the $(\mu_Y,\mu_T)$; see Online Appendix~\ref{section:rkd} for details.

\section{Numerical Illustrations} \label{sec:simulation}
\subsection{Empirical Application}\label{sec::application}

\begin{figure}[!t] 
	\centering	
	\resizebox{0.84\linewidth}{!}{\input{Graphics/applicationmean}}
	\resizebox{0.84\linewidth}{!}{\input{Graphics/applicationfraction}}	
	\caption{Average log consumption  (top panel) and empirical proportion of retired individuals by years of age to/from the retirement eligibility threshold.  Dashed vertical lines indicate indicates the cutoff, which is normalized to zero. Size of  dots is proportional to the respective number of individuals in the data. }\label{applicationsummaryplot}
\end{figure}

In this subsection, we illustrate our methods revisiting data from \citet{battistin2009retirement}, who study  the effects of retirement on consumption in Italy. The data are a sample of $n=30,006$ individuals, obtained by combining several waves of the Bank of Italy Survey on Household Income and Wealth (SHIW) for the period 1993-–2004. We take the natural logarithm of total household spending as the outcome, retirement as the treatment,  and years of age from the formal retirement eligibility threshold, which is normalized to zero, as the running variable. The running variable is thus discrete, but still has somewhat rich support. Figure~\ref{applicationsummaryplot} shows the average of log consumption and the empirical proportion of retired individuals in the data as a function of the running variable.

We then compute our bias-aware AR CSs with both rules of thumb ROT1 and ROT2 to choose the 
smoothness bounds. The resulting CSs turn out to be intervals.
We compare them to bias-aware DM CIs that use the same smoothness bounds, to robust bias correction DM CIs, to DM CIs based on global linear regression with separate intercepts and slopes on each  side of the cutoff, and to DM CIs based on a global 4th order polynomial regression with a dummy  for retirement eligibility.\footnote{In Online Appendix~\ref{appendix::optimized}, we also report results for variants of these CSs that use the optimized RD estimator of \cite{imbens2019optimized} instead of local linear regression.}
We report the results in Table~\ref{table_application} in the form ``midpoint $\pm$ half-length'' to make comparing differences in the CIs' location and length easier. 

\begin{table}[!t]
	\centering
	\begin{threeparttable}	\caption{Confidence sets for the effect of retiring on log consumption for various methods}\label{table_application}
		\begin{tabular}{ccc}
			\toprule
			Smoothness Bound & Method & Confidence Set \\
			\midrule
			\multirow{2}{*}{ROT1 ($B_Y=0.004, B_T= 0.008$)} & Bias-aware AR CS & $-0.268 \pm 0.356$\\
			& Bias-aware DM CS &  $-0.216 \pm 0.304$\\\midrule
			\multirow{2}{*}{ROT2 ($B_Y=0.002, B_T= 0.002$)} & Bias-aware AR CS & $-0.150 \pm 0.260$\\
			& Bias-aware DM CS & $-0.136 \pm 0.234$\\ \midrule
			& Robust bias correction DM CI & $-0.269 \pm 0.305$\\
			-- & Global Linear  with DM CI & $-0.252 \pm 0.065$\\
			& Global  Polynomial with DM CI& $-0.376 \pm 0.042$\\
			\bottomrule 
		\end{tabular}
		\begin{tablenotes}
			\item \textit{Notes:} 
			30,006 data points, CSs with 95\% nominal coverage. 
		\end{tablenotes}
	\end{threeparttable}
\end{table}
We see that bias-aware AR CSs can differ meaningfully from their bias-aware DM CI counterparts in terms of both length and location, even if the same smoothness bounds are used. Our preferred rule ROT2 produces markedly smaller smoothness bounds than ROT1, which is reflected in the  shorter CSs. Robust bias correction yields DM CIs that are qualitatively closer to those obtained under ROT1 by bias-aware methods. The two global parametric methods are the only ones that yield CIs that do not cover zero, but of course this does not account for the model misspecification bias apparent from Figure~\ref{applicationsummaryplot}.

\subsection{Simulations} \label{sec:simulation}
In this subsection, we compare the practical performance of our bias-aware AR CS to that of alternative procedures through simulations. We consider a number of data generating processes calibrated to the data from \citet{battistin2009retirement} with varying curvature of the conditional expectation functions, richness of the running variable’s support, and strength of identification.

\subsubsection*{Data Generating Processes}

We first create three versions of each of the two  CEFs of outcomes and treatment, shown in Figure~\ref{simulationsummaryplot}. Specifically, for $W_i$ equal to either $T_i$ or $Y_i$, we create the $s$th CEF version $\mu_{W,s}(x)$    by fitting a second order spline with four knots on each side of the cutoff, that is,
\begin{align*}
	\mu_{W,s}(x) &=\1{x\geq 0}\left(\beta_{0,W+} + \beta_{1,W+} x +  \sum_{j=1}^4 \beta_{j+1,W+}  \lfloor x-v_{j}\rfloor^2\right) \\
	&\quad +\1{x< 0}\left(\beta_{0,W-} +\beta_{1,W-} x +   \sum_{j=1}^4 \beta_{j+1,W-}  \lfloor x-v_{j}\rfloor^2\right)
\end{align*}
via least squares to the data from \citet{battistin2009retirement}, with $\lfloor \cdot \rfloor = \max(0, \cdot)$. Here our first version uses the knot points $v_j\in\{0, 10, 20, 30\}$; the second version uses the knot points $v_{j} \in\{0, 5, 15, 30\}$, creating greater curvature near the cutoff; while the third version also uses the knot points $v_{j} \in\{0, 5, 15, 30\}$ and additionally fixes  the intercept parameters $\beta_{0,W+}$ and  $\beta_{0,W-}$ to generate very high curvature near the cutoff.\footnote{For the conditional treatment probability, our least squares fits also impose the constraint that the function is increasing, and that it is equal to zero and one below and above the lowest and highest support point, respectively.}
 The magnitude of the jump in the conditional treatment probability, and hence the strength of identification, also decreases across versions.
We then consider all combinations of these CEFs, but for brevity only report results for the four combinations of $\mu_{T,j}$ and $\mu_{Y,k}$ with $(j,k)\in\{(1,1),(2,2),(2,3),(3,2)\}$. See Online Appendix~\ref{appendix::addl_sim} for the remaining results.
In the following we refer to these settings as having either low, moderate or high CEF curvature. 

\begin{figure}[!t] 
	\centering	
	\resizebox{\linewidth}{!}{\input{Graphics/dgp_simulation}}
	\caption{CEFs of outcome (left panel) and of the treatment (right panel) CEFs used in the simulations. Dashed vertical lines indicate indicates the cutoff, which is normalized to zero. Data  from Figure~\ref{applicationsummaryplot} (grey dots) shown for reference. }\label{simulationsummaryplot}
\end{figure}

For each combination of CEFs, we also consider four different distributions for the running variable $X_i$: the ``mildly discrete'' empirical distribution of the data from \citet{battistin2009retirement}; a continuous distribution obtained by adding uniformly distributed  $U(0,1)$ noise to a draw from the empirical distribution; and two more ``coarsely discrete'' distributions obtained by rounding draws from the empirical distribution to the integers $\{\pm 1, \pm(1+d), \pm(1+2d), \dots \}$ for $d\in \{3,6\}$.   
Finally, for any draw of $X_i$ we draw treatment status $T_i$ from a Bernoulli distribution with mean $\mu_{T,j}(X_i)$ and outcome $Y_i$ from a normal distribution with mean $\mu_{T,k}(X_i)$ and variance $\sigma^2(X_i)$, for $(j,k)\in\{1,3\}\times\{1,2,3\}$ and $\sigma^2(x)$  the sampling variance of $Y_i$ among units with running variable equal to the original draw of the running variable from the empirical distribution in the data.

\subsubsection*{Methods}
We study the performance of eight different implementations of AR CSs in our simulations: (i) our bias-aware CS, using the respective true smoothness bounds $B_Y$ and $B_T$; (ii) our bias-aware CS, using twice the true $B_Y$ and $B_T$; (iii) our bias-aware CS, using half the true $B_Y$ and $B_T$; (iv) our bias-aware CS, using ROT1 estimates of $B_Y$ and $B_T$; (v) our bias-aware CS, using ROT2 estimates of $B_Y$ and $B_T$; (vi) a naive CS that ignores bias, using an estimate of the ``pointwise-MSE optimal'' bandwidth \citep[][henceforth IK]{imbens2012optimal}; (vii) an undersmoothing CS, using $n^{-1/20}$ times the estimated IK bandwidth; and (viii) a robust bias correction CS, using local quadratic regression to estimate the bias, and estimated IK bandwidths. In addition, we also consider the performance of eight different DM CIs using the just-mentioned approaches to handling bias.\footnote{
	Computations are carried out with the statistical software \texttt{R}. All bias-aware CSs are computed using our own software, which builds on the package \texttt{RDHonest}. All other CSs are computed using functions from the package \texttt{rdrobust}. A triangular kernel is used in all cases. We note that the  IK bandwidth estimates computed by \texttt{rdrobust} are sometimes too small for the respective CSs to be well-defined  if the running variable is discrete. In those cases, we manually set the main bandwidth such that positive weights are given to three support points on each side of the cutoff (for the bias correction bandwidth we use four support points). In Online Appendix~\ref{appendix::optimized}, we also report results for variants of our CSs in which local linear regression is replaced with the optimized RD estimator of \cite{imbens2019optimized} that are based on the package \texttt{optrdd}.}

\subsubsection*{Results} Table~\ref{tablesimulatedcoveragerates_main} shows the simulated coverage rates of the various CSs under the sixteen different DGPs we consider in our simulations (four combinations of CEFs times four  running variable distributions). We first discuss results for AR CSs, shown in the left panel.
With the true smoothness bounds, the coverage rates of our bias-aware CSs are close to and mostly slightly above the nominal level, irrespective of running variable distribution, curvature of the unknown functions, and identification strength. The slight overcoverage occurs because the function $\mu_Y(x) - \theta\mu_T(x)$ is not exactly quadratic in either setting, and thus the bias does not achieve its worst-case value. Using twice the true bounds increases simulated coverage as expected, while half the true value results in meaningful undercoverage in some settings

Using one of the ROTs for the smoothness bounds leads to potentially severe distortions in some settings, which highlights the need to investigate the fit of the respective underlying global polynomial approximation in practice (cf.\ Online Appendix~\ref{sec:appendixB}).
 Combining a naive approach, undersmoothing, or robust bias correction with an AR construction  leads to CSs with  undercoverage that is modest in some DGPs we consider, but can be substantial especially for those with more coarse running variable support, stronger curvature of the CEFs, and weak identification (high treatment CEF curvature).
  
 Turning to results for DM CIs in the right panel of Table~\ref{tablesimulatedcoveragerates_main}, we see that combining a bias-aware approach with this construction does not lead to CIs with correct coverage in all settings even when using the true smoothness bound. This is because bias-aware DM CIs only control the bias of a first-order approximation of the estimator on which they are based. Coverage distortions are particularly severe in settings of strong curvature, and they further amplify in settings with weak identification. Using the ROT choices for the smoothness bounds leads to further distortions in some cases. 
The coverage of DM CIs that use the naive approach, undersmoothing, or robust bias correction is distorted in most settings,  and the distortions generally become more severe with a more coarse support, in settings with a higher curvature and it further amplifies in settings of weak identification (high treatment CEF curvature). 

\afterpage{	
	\begin{landscape}
		\begin{table}
			\caption{Simulated CS coverage (\%)}\label{tablesimulatedcoveragerates_main}
			\resizebox*{1.35\textwidth}{!}{
				\begin{threeparttable}
					\begin{tabular}{c@{\hskip 0.5in}rrrrrrrr@{\hskip 0.5in}rrrrrrrr}
						\toprule
						& \multicolumn{8}{c}{Anderson-Rubin} &\multicolumn{8}{c}{Delta Method} \\
						\cmidrule[0.4pt](lr{0.3in}){2-9}  \cmidrule[0.4pt](lr{0.1in}){10-17} 	
						& \multicolumn{5}{c}{Bias-Aware} && &&\multicolumn{5}{c}{Bias-Aware} &&&\\
						\cmidrule{2-6}	\cmidrule{10-14} 
						\hspace*{0.75cm} 
						Support	&  TC  & TC$\times .5$ & TC$\times 2$ & ROT1 & ROT2 & Naive & US  & RBC & TC  & TC$\times .5$ & TC$\times 2$ & ROT1 & ROT2 & Naive &   US & RBC  \\ 
						\midrule
						\multicolumn{13}{l}{\textbf{Setting~1} - \textit{Low outcome CEF curvature, low treatment CEF curvature} }&&&&\\  
				Baseline & 96.7 & 91.0 & 98.4 & 97.9 & 92.4 & 88.7 & 93.7 & 94.6 & 97.9 & 93.7 & 99.3 & 98.9 & 93.9 & 94.2 & 96.0 & 94.2 \\ 
				Continuous & 96.6 & 92.7 & 97.4 & 97.1 & 94.3 & 89.6 & 94.0 & 94.7 & 97.2 & 94.4 & 98.0 & 97.8 & 95.1 & 94.7 & 95.4 & 94.7 \\ 
			$\{\pm 1, \pm 4, \dots \}$ & 95.9 & 92.6 & 98.4 & 97.8 & 94.4 & 88.1 & 93.8 & 94.6 & 96.7 & 93.8 & 99.3 & 98.7 & 95.0 & 94.0 & 89.3 & 94.6 \\ 
				$\{\pm 1, \pm 7, \dots \}$  & 97.8 & 91.6 & 100.0 & 99.6 & 95.5 & 78.0 & 84.8 & 93.2 & 98.6 & 92.3 & 100.0 & 99.8 & 96.0 & 85.1 & 77.5 & 93.1 \\ 
				\multicolumn{13}{l}{\phantom{a}}&&&&\\
				\multicolumn{13}{l}{\textbf{Setting~2} - \textit{Moderate outcome CEF curvature, moderate treatment CEF curvature}  }&&&&\\  
						Baseline & 97.2 & 88.7 & 99.2 & 91.9 & 36.4 & 47.6 & 80.9 & 58.6 & 91.1 & 74.0 & 98.3 & 81.9 & 24.7 & 73.3 & 93.0 & 72.9 \\ 
						Continuous & 96.4 & 92.2 & 97.3 & 94.3 & 59.8 & 68.6 & 87.6 & 78.2 & 93.7 & 84.7 & 96.5 & 89.3 & 46.6 & 84.9 & 94.5 & 86.3 \\ 
						$\{\pm 1, \pm 4, \dots \}$ & 98.3 & 88.8 & 100 & 92.1 & 63.5 & 27.5 & 71.1 & 45.6 & 92.7 & 81.8 & 99.0 & 86.0 & 46.6 & 70.7 & 65.9 & 71.3 \\ 
						$\{\pm 1, \pm 7, \dots \}$ & 100 & 79.9 & 100 & 92.1 & 26.2 & 1.3 & 6.5 & 0.3 & 95.4 & 56.0 & 100 & 78.2 & 19.8 & 6.0 & 5.5 & 4.7 \\ 
						\multicolumn{13}{l}{\phantom{a}}&&&&\\
						\multicolumn{13}{l}{\textbf{Setting~3} -  \textit{High outcome CEF curvature, moderate treatment CEF curvature}  }&&&&\\  
						Baseline  & 96.5 & 80.9 & 99.3 & 72.2 & 0.0 & 84.2 & 83.1 & 92.9 & 82.3 & 48.8 & 93.9 & 33.0 & 0.0 & 14.6 & 69.1 & 26.0 \\ 
						Continuous & 95.8 & 90.1 & 97.1 & 90.5 & 1.7 & 89.4 & 94.1 & 94.3 & 89.5 & 75.9 & 92.1 & 77.3 & 0.0 & 48.6 & 85.6 & 66.0 \\ 
						$\{\pm 1, \pm 4, \dots \}$	 & 98.6 & 52.1 & 100 & 55.0 & 6.1 & 4.9 & 5.0 & 84.0 & 78.7 & 25.5 & 100 & 27.9 & 0.0 & 1.9 & 1.9 & 5.9 \\ 
						$\{\pm 1, \pm 7, \dots \}$	 & 100 & 2.7 & 100 & 6.6 & 0.0 & 0.0 & 2.4 & 0.0 & 61.0 & 0.0 & 100 & 0.1 & 0.0 & 0.0 & 0.0 & 0.0 \\
							\multicolumn{13}{l}{\phantom{a}}&&&&\\
						\multicolumn{13}{l}{\textbf{Setting~4} - \textit{Moderate outcome CEF curvature, high treatment CEF curvature}  }&&&&\\  
						Baseline  & 96.8 & 82.1 & 99.4 & 73.0 & 0.0 & 85.9 & 84.2 & 93.4 & 23.5 & 0.1 & 58.5 & 1.2 & 0.0 & 14.2 & 55.0 & 16.6 \\ 
						Continuous & 95.8 & 90.1 & 97.0 & 90.9 & 0.8 & 89.7 & 94.0 & 94.4 & 50.8 & 2.1 & 71.3 & 16.3 & 0.0 & 53.0 & 72.8 & 58.2 \\ 
						$\{\pm 1, \pm 4, \dots \}$	 & 99.0 & 55.9 & 100 & 60.5 & 6.6 & 5.3 & 5.1 & 79.5 & 12.8 & 3.7 & 40.6 & 5.3 & 0.0 & 0.9 & 0.8 & 3.9 \\ 
						$\{\pm 1, \pm 7, \dots \}$	 & 100 & 5.2 & 100 & 17.6 & 0.0 & 0.0 & 2.0 & 0.0 & 0.0 & 0.0 & 3.1 & 0.0 & 0.0 & 0.0 & 1.0 & 0.0 \\ 
						\bottomrule
						\end{tabular}
						\begin{tablenotes}
						\item \textit{Notes:} Results based on 50,000 Monte Carlo draws for a nominal confidence level of 95\%. Columns show results for bias aware approach with true constants (TC), two times true constants (TC$\times2$), half  true constants (TC$\times.5$), and with rule of thumb estimates (ROT1) and (ROT2); naive approach that ignores bias (Naive); undersmoothing  (US);  and robust bias correction (RBC).
						\end{tablenotes}
					\end{threeparttable}}
				\end{table}
			\end{landscape}}

\afterpage{	
	\begin{landscape}
		\begin{figure}[!t]
			\centering
			\begin{minipage}{.65\textwidth}
				\centering (a) Bias-aware Anderson-Rubin CSs 
			
			\vspace{-1cm}
			\resizebox{\linewidth}{!}{	
\begin{tikzpicture}[x=1pt,y=1pt]
\definecolor{fillColor}{RGB}{255,255,255}
\path[use as bounding box,fill=fillColor,fill opacity=0.00] (0,0) rectangle (361.35,252.94);
\begin{scope}
\path[clip] ( 49.20, 61.20) rectangle (336.15,203.75);
\definecolor{drawColor}{RGB}{0,0,0}

\path[draw=drawColor,line width= 0.4pt,line join=round,line cap=round] (-28.74, 66.48) --
	( 15.55, 66.48) --
	( 45.07, 66.49) --
	( 56.42, 66.50) --
	( 67.78, 66.54) --
	( 79.13, 66.76) --
	( 90.48, 67.34) --
	(101.84, 69.08) --
	(113.19, 73.24) --
	(124.55, 82.15) --
	(135.90, 98.39) --
	(147.26,121.36) --
	(158.61,146.96) --
	(169.97,167.83) --
	(181.32,180.15) --
	(192.68,183.50) --
	(204.03,179.99) --
	(215.38,169.25) --
	(226.74,150.77) --
	(238.09,127.19) --
	(249.45,104.25) --
	(260.80, 86.51) --
	(272.16, 75.73) --
	(283.51, 70.24) --
	(294.87, 67.89) --
	(306.22, 66.98) --
	(317.57, 66.65) --
	(328.93, 66.51) --
	(340.28, 66.48) --
	(369.80, 66.48) --
	(414.09, 66.48);
\end{scope}
\begin{scope}
\path[clip] (  0.00,  0.00) rectangle (361.35,252.94);
\definecolor{drawColor}{RGB}{0,0,0}

\node[text=drawColor,anchor=base,inner sep=0pt, outer sep=0pt, scale=  1.00] at (192.68, 15.60) {Distance to theta};

\node[text=drawColor,rotate= 90.00,anchor=base,inner sep=0pt, outer sep=0pt, scale=  1.00] at ( 10.80,132.47) {Simulated Coverage Probability};
\end{scope}
\begin{scope}
\path[clip] ( 49.20, 61.20) rectangle (336.15,203.75);
\definecolor{drawColor}{RGB}{27,158,119}

\path[draw=drawColor,line width= 0.4pt,line join=round,line cap=round] (-28.74, 66.48) --
	( 15.55, 66.48) --
	( 45.07, 66.48) --
	( 56.42, 66.48) --
	( 67.78, 66.48) --
	( 79.13, 66.50) --
	( 90.48, 66.56) --
	(101.84, 66.97) --
	(113.19, 68.62) --
	(124.55, 73.70) --
	(135.90, 86.99) --
	(147.26,111.41) --
	(158.61,142.93) --
	(169.97,169.16) --
	(181.32,182.06) --
	(192.68,182.31) --
	(204.03,172.48) --
	(215.38,151.49) --
	(226.74,123.01) --
	(238.09, 96.42) --
	(249.45, 78.91) --
	(260.80, 70.54) --
	(272.16, 67.64) --
	(283.51, 66.77) --
	(294.87, 66.53) --
	(306.22, 66.48) --
	(317.57, 66.48) --
	(328.93, 66.48) --
	(340.28, 66.48) --
	(369.80, 66.48) --
	(414.09, 66.48);
\definecolor{drawColor}{RGB}{217,95,2}

\path[draw=drawColor,line width= 0.4pt,line join=round,line cap=round] (-28.74, 66.48) --
	( 15.55, 66.49) --
	( 45.07, 66.62) --
	( 56.42, 66.84) --
	( 67.78, 67.29) --
	( 79.13, 68.35) --
	( 90.48, 70.49) --
	(101.84, 74.69) --
	(113.19, 82.63) --
	(124.55, 95.07) --
	(135.90,112.74) --
	(147.26,133.51) --
	(158.61,153.73) --
	(169.97,169.98) --
	(181.32,179.70) --
	(192.68,183.42) --
	(204.03,182.12) --
	(215.38,176.00) --
	(226.74,164.26) --
	(238.09,147.33) --
	(249.45,127.37) --
	(260.80,107.99) --
	(272.16, 92.21) --
	(283.51, 80.82) --
	(294.87, 74.00) --
	(306.22, 70.26) --
	(317.57, 68.25) --
	(328.93, 67.29) --
	(340.28, 66.82) --
	(369.80, 66.52) --
	(414.09, 66.48);
\definecolor{drawColor}{RGB}{117,112,179}

\path[draw=drawColor,line width= 0.4pt,line join=round,line cap=round] (-28.74, 66.48) --
	( 15.55, 66.49) --
	( 45.07, 66.61) --
	( 56.42, 66.78) --
	( 67.78, 67.17) --
	( 79.13, 68.02) --
	( 90.48, 69.87) --
	(101.84, 73.51) --
	(113.19, 80.65) --
	(124.55, 92.72) --
	(135.90,110.36) --
	(147.26,132.18) --
	(158.61,153.30) --
	(169.97,170.12) --
	(181.32,179.85) --
	(192.68,183.26) --
	(204.03,181.55) --
	(215.38,174.59) --
	(226.74,161.59) --
	(238.09,143.87) --
	(249.45,123.88) --
	(260.80,105.34) --
	(272.16, 90.58) --
	(283.51, 80.24) --
	(294.87, 73.85) --
	(306.22, 70.36) --
	(317.57, 68.41) --
	(328.93, 67.42) --
	(340.28, 66.93) --
	(369.80, 66.56) --
	(414.09, 66.48);
\definecolor{drawColor}{RGB}{230,171,2}

\path[draw=drawColor,line width= 0.4pt,line join=round,line cap=round] (-28.74, 66.48) --
	( 15.55, 66.48) --
	( 45.07, 66.48) --
	( 56.42, 66.48) --
	( 67.78, 66.48) --
	( 79.13, 66.50) --
	( 90.48, 66.55) --
	(101.84, 66.96) --
	(113.19, 68.72) --
	(124.55, 74.34) --
	(135.90, 88.61) --
	(147.26,113.44) --
	(158.61,143.93) --
	(169.97,168.30) --
	(181.32,181.00) --
	(192.68,182.68) --
	(204.03,175.50) --
	(215.38,157.91) --
	(226.74,131.67) --
	(238.09,104.21) --
	(249.45, 84.40) --
	(260.80, 73.08) --
	(272.16, 68.56) --
	(283.51, 67.06) --
	(294.87, 66.59) --
	(306.22, 66.50) --
	(317.57, 66.48) --
	(328.93, 66.48) --
	(340.28, 66.48) --
	(369.80, 66.48) --
	(414.09, 66.48);
\end{scope}
\begin{scope}
\path[clip] (  0.00,  0.00) rectangle (361.35,252.94);
\definecolor{drawColor}{RGB}{0,0,0}

\path[draw=drawColor,line width= 0.4pt,line join=round,line cap=round] ( 49.20, 61.20) -- (336.15, 61.20);

\path[draw=drawColor,line width= 0.4pt,line join=round,line cap=round] ( 81.97, 61.20) -- ( 81.97, 55.20);

\path[draw=drawColor,line width= 0.4pt,line join=round,line cap=round] (137.32, 61.20) -- (137.32, 55.20);

\path[draw=drawColor,line width= 0.4pt,line join=round,line cap=round] (192.68, 61.20) -- (192.68, 55.20);

\path[draw=drawColor,line width= 0.4pt,line join=round,line cap=round] (248.03, 61.20) -- (248.03, 55.20);

\path[draw=drawColor,line width= 0.4pt,line join=round,line cap=round] (303.38, 61.20) -- (303.38, 55.20);

\node[text=drawColor,anchor=base,inner sep=0pt, outer sep=0pt, scale=  1.00] at ( 81.97, 39.60) {-0.5};

\node[text=drawColor,anchor=base,inner sep=0pt, outer sep=0pt, scale=  1.00] at (137.32, 39.60) {-0.25};

\node[text=drawColor,anchor=base,inner sep=0pt, outer sep=0pt, scale=  1.00] at (192.68, 39.60) {0};

\node[text=drawColor,anchor=base,inner sep=0pt, outer sep=0pt, scale=  1.00] at (248.03, 39.60) {0.25};

\node[text=drawColor,anchor=base,inner sep=0pt, outer sep=0pt, scale=  1.00] at (303.38, 39.60) {0.5};

\path[draw=drawColor,line width= 0.4pt,line join=round,line cap=round] ( 49.20, 61.20) -- ( 49.20,203.75);

\path[draw=drawColor,line width= 0.4pt,line join=round,line cap=round] ( 49.20, 66.48) -- ( 43.20, 66.48);

\path[draw=drawColor,line width= 0.4pt,line join=round,line cap=round] ( 49.20, 90.48) -- ( 43.20, 90.48);

\path[draw=drawColor,line width= 0.4pt,line join=round,line cap=round] ( 49.20,114.47) -- ( 43.20,114.47);

\path[draw=drawColor,line width= 0.4pt,line join=round,line cap=round] ( 49.20,138.47) -- ( 43.20,138.47);

\path[draw=drawColor,line width= 0.4pt,line join=round,line cap=round] ( 49.20,162.47) -- ( 43.20,162.47);

\path[draw=drawColor,line width= 0.4pt,line join=round,line cap=round] ( 49.20,186.47) -- ( 43.20,186.47);

\node[text=drawColor,anchor=base east,inner sep=0pt, outer sep=0pt, scale=  1.00] at ( 37.20, 63.04) {0};

\node[text=drawColor,anchor=base east,inner sep=0pt, outer sep=0pt, scale=  1.00] at ( 37.20, 87.03) {0.2};

\node[text=drawColor,anchor=base east,inner sep=0pt, outer sep=0pt, scale=  1.00] at ( 37.20,111.03) {0.4};

\node[text=drawColor,anchor=base east,inner sep=0pt, outer sep=0pt, scale=  1.00] at ( 37.20,135.03) {0.6};

\node[text=drawColor,anchor=base east,inner sep=0pt, outer sep=0pt, scale=  1.00] at ( 37.20,159.03) {0.8};

\node[text=drawColor,anchor=base east,inner sep=0pt, outer sep=0pt, scale=  1.00] at ( 37.20,183.02) {1};
\end{scope}
\begin{scope}
\path[clip] ( 49.20, 61.20) rectangle (336.15,203.75);
\definecolor{drawColor}{RGB}{169,169,169}

\path[draw=drawColor,line width= 0.4pt,dash pattern=on 4pt off 4pt ,line join=round,line cap=round] ( 49.20,180.47) -- (336.15,180.47);

\path[draw=drawColor,line width= 0.4pt,dash pattern=on 4pt off 4pt ,line join=round,line cap=round] (200.00, 61.20) -- (200.00,203.75);
\definecolor{drawColor}{RGB}{0,0,0}

\path[draw=drawColor,line width= 0.8pt,line join=round,line cap=round] (267.20,169.67) -- (283.40,169.67);
\definecolor{drawColor}{RGB}{27,158,119}

\path[draw=drawColor,line width= 0.8pt,line join=round,line cap=round] (267.20,158.87) -- (283.40,158.87);
\definecolor{drawColor}{RGB}{217,95,2}

\path[draw=drawColor,line width= 0.8pt,line join=round,line cap=round] (267.20,148.07) -- (283.40,148.07);
\definecolor{drawColor}{RGB}{117,112,179}

\path[draw=drawColor,line width= 0.8pt,line join=round,line cap=round] (267.20,137.27) -- (283.40,137.27);
\definecolor{drawColor}{RGB}{230,171,2}

\path[draw=drawColor,line width= 0.8pt,line join=round,line cap=round] (267.20,126.47) -- (283.40,126.47);
\definecolor{drawColor}{RGB}{0,0,0}

\node[text=drawColor,anchor=base west,inner sep=0pt, outer sep=0pt, scale=  0.90] at (291.50,166.57) {TC-AR};

\node[text=drawColor,anchor=base west,inner sep=0pt, outer sep=0pt, scale=  0.90] at (291.50,155.77) {TC x 0.5};

\node[text=drawColor,anchor=base west,inner sep=0pt, outer sep=0pt, scale=  0.90] at (291.50,144.97) {TC x 2};

\node[text=drawColor,anchor=base west,inner sep=0pt, outer sep=0pt, scale=  0.90] at (291.50,134.17) {ROT 1};

\node[text=drawColor,anchor=base west,inner sep=0pt, outer sep=0pt, scale=  0.90] at (291.50,123.37) {ROT 2};
\end{scope}
\end{tikzpicture}}					
			\end{minipage}\hfill	
			\begin{minipage}{.65\textwidth}
				\centering (b) Other Anderson-Rubin CSs
				
				\vspace{-1cm}
				\resizebox{\linewidth}{!}{
\begin{tikzpicture}[x=1pt,y=1pt]
\definecolor{fillColor}{RGB}{255,255,255}
\path[use as bounding box,fill=fillColor,fill opacity=0.00] (0,0) rectangle (361.35,252.94);
\begin{scope}
\path[clip] ( 49.20, 61.20) rectangle (336.15,203.75);
\definecolor{drawColor}{RGB}{0,0,0}

\path[draw=drawColor,line width= 0.4pt,line join=round,line cap=round] (-28.74, 66.48) --
	( 15.55, 66.48) --
	( 45.07, 66.49) --
	( 56.42, 66.50) --
	( 67.78, 66.54) --
	( 79.13, 66.76) --
	( 90.48, 67.34) --
	(101.84, 69.08) --
	(113.19, 73.24) --
	(124.55, 82.15) --
	(135.90, 98.39) --
	(147.26,121.36) --
	(158.61,146.96) --
	(169.97,167.83) --
	(181.32,180.15) --
	(192.68,183.50) --
	(204.03,179.99) --
	(215.38,169.25) --
	(226.74,150.77) --
	(238.09,127.19) --
	(249.45,104.25) --
	(260.80, 86.51) --
	(272.16, 75.73) --
	(283.51, 70.24) --
	(294.87, 67.89) --
	(306.22, 66.98) --
	(317.57, 66.65) --
	(328.93, 66.51) --
	(340.28, 66.48) --
	(369.80, 66.48) --
	(414.09, 66.48);
\end{scope}
\begin{scope}
\path[clip] (  0.00,  0.00) rectangle (361.35,252.94);
\definecolor{drawColor}{RGB}{0,0,0}

\node[text=drawColor,anchor=base,inner sep=0pt, outer sep=0pt, scale=  1.00] at (192.68, 15.60) {Parameter Value};

\node[text=drawColor,rotate= 90.00,anchor=base,inner sep=0pt, outer sep=0pt, scale=  1.00] at ( 10.80,132.47) {Simulated Coverage Probability};
\end{scope}
\begin{scope}
\path[clip] ( 49.20, 61.20) rectangle (336.15,203.75);
\definecolor{drawColor}{RGB}{66,134,244}

\path[draw=drawColor,line width= 0.4pt,line join=round,line cap=round] (-28.74, 66.48) --
	( 15.55, 66.48) --
	( 45.07, 66.49) --
	( 56.42, 66.49) --
	( 67.78, 66.50) --
	( 79.13, 66.52) --
	( 90.48, 66.57) --
	(101.84, 66.73) --
	(113.19, 67.07) --
	(124.55, 68.47) --
	(135.90, 74.95) --
	(147.26, 95.17) --
	(158.61,127.91) --
	(169.97,157.34) --
	(181.32,174.18) --
	(192.68,177.51) --
	(204.03,169.71) --
	(215.38,153.04) --
	(226.74,131.75) --
	(238.09,110.47) --
	(249.45, 93.07) --
	(260.80, 80.90) --
	(272.16, 73.85) --
	(283.51, 70.08) --
	(294.87, 68.25) --
	(306.22, 67.33) --
	(317.57, 66.92) --
	(328.93, 66.69) --
	(340.28, 66.58) --
	(369.80, 66.51) --
	(414.09, 66.48);
\definecolor{drawColor}{RGB}{102,166,30}

\path[draw=drawColor,line width= 0.4pt,line join=round,line cap=round] (-28.74, 66.50) --
	( 15.55, 66.53) --
	( 45.07, 66.66) --
	( 56.42, 66.76) --
	( 67.78, 66.91) --
	( 79.13, 67.19) --
	( 90.48, 67.58) --
	(101.84, 68.29) --
	(113.19, 70.13) --
	(124.55, 75.67) --
	(135.90, 90.27) --
	(147.26,114.56) --
	(158.61,141.84) --
	(169.97,163.01) --
	(181.32,175.34) --
	(192.68,179.75) --
	(204.03,177.54) --
	(215.38,169.52) --
	(226.74,156.97) --
	(238.09,141.74) --
	(249.45,125.58) --
	(260.80,110.36) --
	(272.16, 97.45) --
	(283.51, 87.32) --
	(294.87, 79.92) --
	(306.22, 75.08) --
	(317.57, 71.83) --
	(328.93, 69.84) --
	(340.28, 68.63) --
	(369.80, 67.13) --
	(414.09, 66.62);
\definecolor{drawColor}{RGB}{166,118,29}

\path[draw=drawColor,line width= 0.4pt,line join=round,line cap=round] (-28.74, 66.48) --
	( 15.55, 66.48) --
	( 45.07, 66.50) --
	( 56.42, 66.52) --
	( 67.78, 66.57) --
	( 79.13, 66.69) --
	( 90.48, 66.87) --
	(101.84, 67.25) --
	(113.19, 68.10) --
	(124.55, 70.19) --
	(135.90, 76.32) --
	(147.26, 93.00) --
	(158.61,120.55) --
	(169.97,149.75) --
	(181.32,169.76) --
	(192.68,179.00) --
	(204.03,178.70) --
	(215.38,170.16) --
	(226.74,154.44) --
	(238.09,134.77) --
	(249.45,114.27) --
	(260.80, 97.32) --
	(272.16, 84.70) --
	(283.51, 76.69) --
	(294.87, 71.92) --
	(306.22, 69.43) --
	(317.57, 68.02) --
	(328.93, 67.31) --
	(340.28, 66.93) --
	(369.80, 66.57) --
	(414.09, 66.49);
\end{scope}
\begin{scope}
\path[clip] (  0.00,  0.00) rectangle (361.35,252.94);
\definecolor{drawColor}{RGB}{0,0,0}

\path[draw=drawColor,line width= 0.4pt,line join=round,line cap=round] ( 49.20, 61.20) -- (336.15, 61.20);

\path[draw=drawColor,line width= 0.4pt,line join=round,line cap=round] ( 81.97, 61.20) -- ( 81.97, 55.20);

\path[draw=drawColor,line width= 0.4pt,line join=round,line cap=round] (137.32, 61.20) -- (137.32, 55.20);

\path[draw=drawColor,line width= 0.4pt,line join=round,line cap=round] (192.68, 61.20) -- (192.68, 55.20);

\path[draw=drawColor,line width= 0.4pt,line join=round,line cap=round] (248.03, 61.20) -- (248.03, 55.20);

\path[draw=drawColor,line width= 0.4pt,line join=round,line cap=round] (303.38, 61.20) -- (303.38, 55.20);

\node[text=drawColor,anchor=base,inner sep=0pt, outer sep=0pt, scale=  1.00] at ( 81.97, 39.60) {-0.5};

\node[text=drawColor,anchor=base,inner sep=0pt, outer sep=0pt, scale=  1.00] at (137.32, 39.60) {-0.25};

\node[text=drawColor,anchor=base,inner sep=0pt, outer sep=0pt, scale=  1.00] at (192.68, 39.60) {0};

\node[text=drawColor,anchor=base,inner sep=0pt, outer sep=0pt, scale=  1.00] at (248.03, 39.60) {0.25};

\node[text=drawColor,anchor=base,inner sep=0pt, outer sep=0pt, scale=  1.00] at (303.38, 39.60) {0.5};

\path[draw=drawColor,line width= 0.4pt,line join=round,line cap=round] ( 49.20, 61.20) -- ( 49.20,203.75);

\path[draw=drawColor,line width= 0.4pt,line join=round,line cap=round] ( 49.20, 66.48) -- ( 43.20, 66.48);

\path[draw=drawColor,line width= 0.4pt,line join=round,line cap=round] ( 49.20, 90.48) -- ( 43.20, 90.48);

\path[draw=drawColor,line width= 0.4pt,line join=round,line cap=round] ( 49.20,114.47) -- ( 43.20,114.47);

\path[draw=drawColor,line width= 0.4pt,line join=round,line cap=round] ( 49.20,138.47) -- ( 43.20,138.47);

\path[draw=drawColor,line width= 0.4pt,line join=round,line cap=round] ( 49.20,162.47) -- ( 43.20,162.47);

\path[draw=drawColor,line width= 0.4pt,line join=round,line cap=round] ( 49.20,186.47) -- ( 43.20,186.47);

\node[text=drawColor,anchor=base east,inner sep=0pt, outer sep=0pt, scale=  1.00] at ( 37.20, 63.04) {0};

\node[text=drawColor,anchor=base east,inner sep=0pt, outer sep=0pt, scale=  1.00] at ( 37.20, 87.03) {0.2};

\node[text=drawColor,anchor=base east,inner sep=0pt, outer sep=0pt, scale=  1.00] at ( 37.20,111.03) {0.4};

\node[text=drawColor,anchor=base east,inner sep=0pt, outer sep=0pt, scale=  1.00] at ( 37.20,135.03) {0.6};

\node[text=drawColor,anchor=base east,inner sep=0pt, outer sep=0pt, scale=  1.00] at ( 37.20,159.03) {0.8};

\node[text=drawColor,anchor=base east,inner sep=0pt, outer sep=0pt, scale=  1.00] at ( 37.20,183.02) {1};
\end{scope}
\begin{scope}
\path[clip] ( 49.20, 61.20) rectangle (336.15,203.75);
\definecolor{drawColor}{RGB}{169,169,169}

\path[draw=drawColor,line width= 0.4pt,dash pattern=on 4pt off 4pt ,line join=round,line cap=round] ( 49.20,180.47) -- (336.15,180.47);

\path[draw=drawColor,line width= 0.4pt,dash pattern=on 4pt off 4pt ,line join=round,line cap=round] (200.00, 61.20) -- (200.00,203.75);
\definecolor{drawColor}{RGB}{0,0,0}

\path[draw=drawColor,line width= 0.8pt,line join=round,line cap=round] (267.20,169.67) -- (283.40,169.67);
\definecolor{drawColor}{RGB}{66,134,244}

\path[draw=drawColor,line width= 0.8pt,line join=round,line cap=round] (267.20,158.87) -- (283.40,158.87);
\definecolor{drawColor}{RGB}{102,166,30}

\path[draw=drawColor,line width= 0.8pt,line join=round,line cap=round] (267.20,148.07) -- (283.40,148.07);
\definecolor{drawColor}{RGB}{166,118,29}

\path[draw=drawColor,line width= 0.8pt,line join=round,line cap=round] (267.20,137.27) -- (283.40,137.27);
\definecolor{drawColor}{RGB}{0,0,0}

\node[text=drawColor,anchor=base west,inner sep=0pt, outer sep=0pt, scale=  0.90] at (291.50,166.57) {TC-AR};

\node[text=drawColor,anchor=base west,inner sep=0pt, outer sep=0pt, scale=  0.90] at (291.50,155.77) {Naive};

\node[text=drawColor,anchor=base west,inner sep=0pt, outer sep=0pt, scale=  0.90] at (291.50,144.97) {US};

\node[text=drawColor,anchor=base west,inner sep=0pt, outer sep=0pt, scale=  0.90] at (291.50,134.17) {RBC};
\end{scope}
\end{tikzpicture}}
			\end{minipage}\bigskip	
			
			\begin{minipage}{.65\textwidth}
				\centering (c) Bias-Aware  Delta Method CI
			
			\vspace{-1cm}
				\resizebox{\linewidth}{!}{
\begin{tikzpicture}[x=1pt,y=1pt]
\definecolor{fillColor}{RGB}{255,255,255}
\path[use as bounding box,fill=fillColor,fill opacity=0.00] (0,0) rectangle (361.35,252.94);
\begin{scope}
\path[clip] ( 49.20, 61.20) rectangle (336.15,203.75);
\definecolor{drawColor}{RGB}{0,0,0}

\path[draw=drawColor,line width= 0.4pt,line join=round,line cap=round] (-28.74, 66.48) --
	( 15.55, 66.48) --
	( 45.07, 66.49) --
	( 56.42, 66.50) --
	( 67.78, 66.54) --
	( 79.13, 66.76) --
	( 90.48, 67.34) --
	(101.84, 69.08) --
	(113.19, 73.24) --
	(124.55, 82.15) --
	(135.90, 98.39) --
	(147.26,121.36) --
	(158.61,146.96) --
	(169.97,167.83) --
	(181.32,180.15) --
	(192.68,183.50) --
	(204.03,179.99) --
	(215.38,169.25) --
	(226.74,150.77) --
	(238.09,127.19) --
	(249.45,104.25) --
	(260.80, 86.51) --
	(272.16, 75.73) --
	(283.51, 70.24) --
	(294.87, 67.89) --
	(306.22, 66.98) --
	(317.57, 66.65) --
	(328.93, 66.51) --
	(340.28, 66.48) --
	(369.80, 66.48) --
	(414.09, 66.48);
\end{scope}
\begin{scope}
\path[clip] (  0.00,  0.00) rectangle (361.35,252.94);
\definecolor{drawColor}{RGB}{0,0,0}

\node[text=drawColor,anchor=base,inner sep=0pt, outer sep=0pt, scale=  1.00] at (192.68, 15.60) {Parameter Value};

\node[text=drawColor,rotate= 90.00,anchor=base,inner sep=0pt, outer sep=0pt, scale=  1.00] at ( 10.80,132.47) {Simulated Coverage Probability};
\end{scope}
\begin{scope}
\path[clip] ( 49.20, 61.20) rectangle (336.15,203.75);
\definecolor{drawColor}{gray}{0.40}

\path[draw=drawColor,line width= 0.4pt,line join=round,line cap=round] (-28.74, 66.48) --
	( 15.55, 66.48) --
	( 45.07, 66.48) --
	( 56.42, 66.48) --
	( 67.78, 66.50) --
	( 79.13, 66.54) --
	( 90.48, 66.79) --
	(101.84, 67.72) --
	(113.19, 70.43) --
	(124.55, 77.17) --
	(135.90, 91.96) --
	(147.26,115.47) --
	(158.61,144.99) --
	(169.97,168.26) --
	(181.32,180.54) --
	(192.68,183.98) --
	(204.03,180.99) --
	(215.38,169.06) --
	(226.74,144.51) --
	(238.09,114.60) --
	(249.45, 92.46) --
	(260.80, 77.95) --
	(272.16, 70.77) --
	(283.51, 67.92) --
	(294.87, 66.91) --
	(306.22, 66.58) --
	(317.57, 66.49) --
	(328.93, 66.48) --
	(340.28, 66.48) --
	(369.80, 66.48) --
	(414.09, 66.48);
\definecolor{drawColor}{RGB}{27,158,119}

\path[draw=drawColor,line width= 0.4pt,line join=round,line cap=round] (-28.74, 66.48) --
	( 15.55, 66.48) --
	( 45.07, 66.48) --
	( 56.42, 66.48) --
	( 67.78, 66.48) --
	( 79.13, 66.48) --
	( 90.48, 66.51) --
	(101.84, 66.69) --
	(113.19, 67.65) --
	(124.55, 71.41) --
	(135.90, 82.86) --
	(147.26,107.69) --
	(158.61,142.14) --
	(169.97,169.89) --
	(181.32,182.28) --
	(192.68,183.50) --
	(204.03,174.52) --
	(215.38,148.31) --
	(226.74,108.47) --
	(238.09, 84.72) --
	(249.45, 72.24) --
	(260.80, 67.92) --
	(272.16, 66.76) --
	(283.51, 66.50) --
	(294.87, 66.48) --
	(306.22, 66.48) --
	(317.57, 66.48) --
	(328.93, 66.48) --
	(340.28, 66.48) --
	(369.80, 66.48) --
	(414.09, 66.48);
\definecolor{drawColor}{RGB}{217,95,2}

\path[draw=drawColor,line width= 0.4pt,line join=round,line cap=round] (-28.74, 66.48) --
	( 15.55, 66.48) --
	( 45.07, 66.50) --
	( 56.42, 66.54) --
	( 67.78, 66.67) --
	( 79.13, 67.09) --
	( 90.48, 68.19) --
	(101.84, 70.83) --
	(113.19, 76.28) --
	(124.55, 86.68) --
	(135.90,103.76) --
	(147.26,127.03) --
	(158.61,152.16) --
	(169.97,170.75) --
	(181.32,180.68) --
	(192.68,184.12) --
	(204.03,182.99) --
	(215.38,176.55) --
	(226.74,162.06) --
	(238.09,138.86) --
	(249.45,115.09) --
	(260.80, 95.67) --
	(272.16, 81.87) --
	(283.51, 73.69) --
	(294.87, 69.64) --
	(306.22, 67.73) --
	(317.57, 66.95) --
	(328.93, 66.64) --
	(340.28, 66.53) --
	(369.80, 66.48) --
	(414.09, 66.48);
\definecolor{drawColor}{RGB}{117,112,179}

\path[draw=drawColor,line width= 0.4pt,line join=round,line cap=round] (-28.74, 66.48) --
	( 15.55, 66.48) --
	( 45.07, 66.50) --
	( 56.42, 66.55) --
	( 67.78, 66.67) --
	( 79.13, 66.99) --
	( 90.48, 67.96) --
	(101.84, 70.16) --
	(113.19, 75.00) --
	(124.55, 84.75) --
	(135.90,101.71) --
	(147.26,125.53) --
	(158.61,151.67) --
	(169.97,170.79) --
	(181.32,180.74) --
	(192.68,184.04) --
	(204.03,182.60) --
	(215.38,175.03) --
	(226.74,158.44) --
	(238.09,134.66) --
	(249.45,112.25) --
	(260.80, 94.49) --
	(272.16, 82.05) --
	(283.51, 74.11) --
	(294.87, 70.08) --
	(306.22, 68.13) --
	(317.57, 67.15) --
	(328.93, 66.76) --
	(340.28, 66.59) --
	(369.80, 66.48) --
	(414.09, 66.48);
\definecolor{drawColor}{RGB}{230,171,2}

\path[draw=drawColor,line width= 0.4pt,line join=round,line cap=round] (-28.74, 66.48) --
	( 15.55, 66.48) --
	( 45.07, 66.48) --
	( 56.42, 66.48) --
	( 67.78, 66.48) --
	( 79.13, 66.49) --
	( 90.48, 66.52) --
	(101.84, 66.72) --
	(113.19, 67.93) --
	(124.55, 72.42) --
	(135.90, 85.35) --
	(147.26,110.45) --
	(158.61,142.83) --
	(169.97,168.50) --
	(181.32,181.24) --
	(192.68,183.37) --
	(204.03,176.50) --
	(215.38,156.93) --
	(226.74,125.59) --
	(238.09, 97.34) --
	(249.45, 79.38) --
	(260.80, 70.56) --
	(272.16, 67.60) --
	(283.51, 66.74) --
	(294.87, 66.52) --
	(306.22, 66.48) --
	(317.57, 66.48) --
	(328.93, 66.48) --
	(340.28, 66.48) --
	(369.80, 66.48) --
	(414.09, 66.48);
\end{scope}
\begin{scope}
\path[clip] (  0.00,  0.00) rectangle (361.35,252.94);
\definecolor{drawColor}{RGB}{0,0,0}

\path[draw=drawColor,line width= 0.4pt,line join=round,line cap=round] ( 49.20, 61.20) -- (336.15, 61.20);

\path[draw=drawColor,line width= 0.4pt,line join=round,line cap=round] ( 81.97, 61.20) -- ( 81.97, 55.20);

\path[draw=drawColor,line width= 0.4pt,line join=round,line cap=round] (137.32, 61.20) -- (137.32, 55.20);

\path[draw=drawColor,line width= 0.4pt,line join=round,line cap=round] (192.68, 61.20) -- (192.68, 55.20);

\path[draw=drawColor,line width= 0.4pt,line join=round,line cap=round] (248.03, 61.20) -- (248.03, 55.20);

\path[draw=drawColor,line width= 0.4pt,line join=round,line cap=round] (303.38, 61.20) -- (303.38, 55.20);

\node[text=drawColor,anchor=base,inner sep=0pt, outer sep=0pt, scale=  1.00] at ( 81.97, 39.60) {-0.5};

\node[text=drawColor,anchor=base,inner sep=0pt, outer sep=0pt, scale=  1.00] at (137.32, 39.60) {-0.25};

\node[text=drawColor,anchor=base,inner sep=0pt, outer sep=0pt, scale=  1.00] at (192.68, 39.60) {0};

\node[text=drawColor,anchor=base,inner sep=0pt, outer sep=0pt, scale=  1.00] at (248.03, 39.60) {0.25};

\node[text=drawColor,anchor=base,inner sep=0pt, outer sep=0pt, scale=  1.00] at (303.38, 39.60) {0.5};

\path[draw=drawColor,line width= 0.4pt,line join=round,line cap=round] ( 49.20, 61.20) -- ( 49.20,203.75);

\path[draw=drawColor,line width= 0.4pt,line join=round,line cap=round] ( 49.20, 66.48) -- ( 43.20, 66.48);

\path[draw=drawColor,line width= 0.4pt,line join=round,line cap=round] ( 49.20, 90.48) -- ( 43.20, 90.48);

\path[draw=drawColor,line width= 0.4pt,line join=round,line cap=round] ( 49.20,114.47) -- ( 43.20,114.47);

\path[draw=drawColor,line width= 0.4pt,line join=round,line cap=round] ( 49.20,138.47) -- ( 43.20,138.47);

\path[draw=drawColor,line width= 0.4pt,line join=round,line cap=round] ( 49.20,162.47) -- ( 43.20,162.47);

\path[draw=drawColor,line width= 0.4pt,line join=round,line cap=round] ( 49.20,186.47) -- ( 43.20,186.47);

\node[text=drawColor,anchor=base east,inner sep=0pt, outer sep=0pt, scale=  1.00] at ( 37.20, 63.04) {0};

\node[text=drawColor,anchor=base east,inner sep=0pt, outer sep=0pt, scale=  1.00] at ( 37.20, 87.03) {0.2};

\node[text=drawColor,anchor=base east,inner sep=0pt, outer sep=0pt, scale=  1.00] at ( 37.20,111.03) {0.4};

\node[text=drawColor,anchor=base east,inner sep=0pt, outer sep=0pt, scale=  1.00] at ( 37.20,135.03) {0.6};

\node[text=drawColor,anchor=base east,inner sep=0pt, outer sep=0pt, scale=  1.00] at ( 37.20,159.03) {0.8};

\node[text=drawColor,anchor=base east,inner sep=0pt, outer sep=0pt, scale=  1.00] at ( 37.20,183.02) {1};
\end{scope}
\begin{scope}
\path[clip] ( 49.20, 61.20) rectangle (336.15,203.75);
\definecolor{drawColor}{RGB}{169,169,169}

\path[draw=drawColor,line width= 0.4pt,dash pattern=on 4pt off 4pt ,line join=round,line cap=round] ( 49.20,180.47) -- (336.15,180.47);

\path[draw=drawColor,line width= 0.4pt,dash pattern=on 4pt off 4pt ,line join=round,line cap=round] (200.00, 61.20) -- (200.00,203.75);
\definecolor{drawColor}{RGB}{0,0,0}

\path[draw=drawColor,line width= 0.8pt,line join=round,line cap=round] (267.20,169.67) -- (283.40,169.67);
\definecolor{drawColor}{gray}{0.40}

\path[draw=drawColor,line width= 0.8pt,line join=round,line cap=round] (267.20,158.87) -- (283.40,158.87);
\definecolor{drawColor}{RGB}{27,158,119}

\path[draw=drawColor,line width= 0.8pt,line join=round,line cap=round] (267.20,148.07) -- (283.40,148.07);
\definecolor{drawColor}{RGB}{217,95,2}

\path[draw=drawColor,line width= 0.8pt,line join=round,line cap=round] (267.20,137.27) -- (283.40,137.27);
\definecolor{drawColor}{RGB}{117,112,179}

\path[draw=drawColor,line width= 0.8pt,line join=round,line cap=round] (267.20,126.47) -- (283.40,126.47);
\definecolor{drawColor}{RGB}{230,171,2}

\path[draw=drawColor,line width= 0.8pt,line join=round,line cap=round] (267.20,115.67) -- (283.40,115.67);
\definecolor{drawColor}{RGB}{0,0,0}

\node[text=drawColor,anchor=base west,inner sep=0pt, outer sep=0pt, scale=  0.90] at (291.50,166.57) {TC-AR};

\node[text=drawColor,anchor=base west,inner sep=0pt, outer sep=0pt, scale=  0.90] at (291.50,155.77) {TC-DM};

\node[text=drawColor,anchor=base west,inner sep=0pt, outer sep=0pt, scale=  0.90] at (291.50,144.97) {TC x 0.5};

\node[text=drawColor,anchor=base west,inner sep=0pt, outer sep=0pt, scale=  0.90] at (291.50,134.17) {TC x 2};

\node[text=drawColor,anchor=base west,inner sep=0pt, outer sep=0pt, scale=  0.90] at (291.50,123.37) {ROT 1};

\node[text=drawColor,anchor=base west,inner sep=0pt, outer sep=0pt, scale=  0.90] at (291.50,112.57) {ROT 2};
\end{scope}
\end{tikzpicture}}		
			\end{minipage}\hfill	
			\begin{minipage}{.65\textwidth}
				\centering (d)  Other Delta Method  CIs
				
				\vspace{-1cm}
				\resizebox{\linewidth}{!}{
\begin{tikzpicture}[x=1pt,y=1pt]
\definecolor{fillColor}{RGB}{255,255,255}
\path[use as bounding box,fill=fillColor,fill opacity=0.00] (0,0) rectangle (361.35,252.94);
\begin{scope}
\path[clip] ( 49.20, 61.20) rectangle (336.15,203.75);
\definecolor{drawColor}{gray}{0.40}

\path[draw=drawColor,line width= 0.4pt,line join=round,line cap=round] (-28.74, 66.48) --
	( 15.55, 66.48) --
	( 45.07, 66.48) --
	( 56.42, 66.48) --
	( 67.78, 66.50) --
	( 79.13, 66.54) --
	( 90.48, 66.79) --
	(101.84, 67.72) --
	(113.19, 70.43) --
	(124.55, 77.17) --
	(135.90, 91.96) --
	(147.26,115.47) --
	(158.61,144.99) --
	(169.97,168.26) --
	(181.32,180.54) --
	(192.68,183.98) --
	(204.03,180.99) --
	(215.38,169.06) --
	(226.74,144.51) --
	(238.09,114.60) --
	(249.45, 92.46) --
	(260.80, 77.95) --
	(272.16, 70.77) --
	(283.51, 67.92) --
	(294.87, 66.91) --
	(306.22, 66.58) --
	(317.57, 66.49) --
	(328.93, 66.48) --
	(340.28, 66.48) --
	(369.80, 66.48) --
	(414.09, 66.48);
\end{scope}
\begin{scope}
\path[clip] (  0.00,  0.00) rectangle (361.35,252.94);
\definecolor{drawColor}{RGB}{0,0,0}

\node[text=drawColor,anchor=base,inner sep=0pt, outer sep=0pt, scale=  1.00] at (192.68, 15.60) {Parameter Value};

\node[text=drawColor,rotate= 90.00,anchor=base,inner sep=0pt, outer sep=0pt, scale=  1.00] at ( 10.80,132.47) {Simulated Coverage Probability};
\end{scope}
\begin{scope}
\path[clip] ( 49.20, 61.20) rectangle (336.15,203.75);
\definecolor{drawColor}{RGB}{66,134,244}

\path[draw=drawColor,line width= 0.4pt,line join=round,line cap=round] (-28.74, 66.48) --
	( 15.55, 66.48) --
	( 45.07, 66.49) --
	( 56.42, 66.50) --
	( 67.78, 66.54) --
	( 79.13, 66.67) --
	( 90.48, 67.04) --
	(101.84, 68.20) --
	(113.19, 71.32) --
	(124.55, 78.10) --
	(135.90, 91.32) --
	(147.26,112.22) --
	(158.61,137.25) --
	(169.97,159.62) --
	(181.32,174.78) --
	(192.68,180.55) --
	(204.03,177.79) --
	(215.38,166.58) --
	(226.74,147.01) --
	(238.09,122.99) --
	(249.45,100.26) --
	(260.80, 83.72) --
	(272.16, 74.16) --
	(283.51, 69.73) --
	(294.87, 67.74) --
	(306.22, 66.94) --
	(317.57, 66.63) --
	(328.93, 66.53) --
	(340.28, 66.49) --
	(369.80, 66.48) --
	(414.09, 66.48);
\definecolor{drawColor}{RGB}{102,166,30}

\path[draw=drawColor,line width= 0.4pt,line join=round,line cap=round] (-28.74, 66.49) --
	( 15.55, 66.55) --
	( 45.07, 66.95) --
	( 56.42, 67.43) --
	( 67.78, 68.28) --
	( 79.13, 70.05) --
	( 90.48, 73.20) --
	(101.84, 78.75) --
	(113.19, 87.44) --
	(124.55,100.30) --
	(135.90,116.51) --
	(147.26,134.83) --
	(158.61,152.29) --
	(169.97,166.73) --
	(181.32,176.18) --
	(192.68,180.66) --
	(204.03,180.44) --
	(215.38,175.50) --
	(226.74,165.78) --
	(238.09,151.64) --
	(249.45,134.33) --
	(260.80,116.28) --
	(272.16,100.47) --
	(283.51, 87.80) --
	(294.87, 79.23) --
	(306.22, 73.55) --
	(317.57, 70.36) --
	(328.93, 68.61) --
	(340.28, 67.58) --
	(369.80, 66.67) --
	(414.09, 66.50);
\definecolor{drawColor}{RGB}{166,118,29}

\path[draw=drawColor,line width= 0.4pt,line join=round,line cap=round] (-28.74, 66.48) --
	( 15.55, 66.48) --
	( 45.07, 66.50) --
	( 56.42, 66.53) --
	( 67.78, 66.60) --
	( 79.13, 66.90) --
	( 90.48, 67.57) --
	(101.84, 69.26) --
	(113.19, 72.91) --
	(124.55, 80.30) --
	(135.90, 93.31) --
	(147.26,112.16) --
	(158.61,134.36) --
	(169.97,155.25) --
	(181.32,170.59) --
	(192.68,178.72) --
	(204.03,179.13) --
	(215.38,172.41) --
	(226.74,158.36) --
	(238.09,138.58) --
	(249.45,116.72) --
	(260.80, 97.26) --
	(272.16, 83.15) --
	(283.51, 74.49) --
	(294.87, 70.17) --
	(306.22, 68.05) --
	(317.57, 67.13) --
	(328.93, 66.74) --
	(340.28, 66.56) --
	(369.80, 66.48) --
	(414.09, 66.48);
\end{scope}
\begin{scope}
\path[clip] (  0.00,  0.00) rectangle (361.35,252.94);
\definecolor{drawColor}{RGB}{0,0,0}

\path[draw=drawColor,line width= 0.4pt,line join=round,line cap=round] ( 49.20, 61.20) -- (336.15, 61.20);

\path[draw=drawColor,line width= 0.4pt,line join=round,line cap=round] ( 81.97, 61.20) -- ( 81.97, 55.20);

\path[draw=drawColor,line width= 0.4pt,line join=round,line cap=round] (137.32, 61.20) -- (137.32, 55.20);

\path[draw=drawColor,line width= 0.4pt,line join=round,line cap=round] (192.68, 61.20) -- (192.68, 55.20);

\path[draw=drawColor,line width= 0.4pt,line join=round,line cap=round] (248.03, 61.20) -- (248.03, 55.20);

\path[draw=drawColor,line width= 0.4pt,line join=round,line cap=round] (303.38, 61.20) -- (303.38, 55.20);

\node[text=drawColor,anchor=base,inner sep=0pt, outer sep=0pt, scale=  1.00] at ( 81.97, 39.60) {-0.5};

\node[text=drawColor,anchor=base,inner sep=0pt, outer sep=0pt, scale=  1.00] at (137.32, 39.60) {-0.25};

\node[text=drawColor,anchor=base,inner sep=0pt, outer sep=0pt, scale=  1.00] at (192.68, 39.60) {0};

\node[text=drawColor,anchor=base,inner sep=0pt, outer sep=0pt, scale=  1.00] at (248.03, 39.60) {0.25};

\node[text=drawColor,anchor=base,inner sep=0pt, outer sep=0pt, scale=  1.00] at (303.38, 39.60) {0.5};

\path[draw=drawColor,line width= 0.4pt,line join=round,line cap=round] ( 49.20, 61.20) -- ( 49.20,203.75);

\path[draw=drawColor,line width= 0.4pt,line join=round,line cap=round] ( 49.20, 66.48) -- ( 43.20, 66.48);

\path[draw=drawColor,line width= 0.4pt,line join=round,line cap=round] ( 49.20, 90.48) -- ( 43.20, 90.48);

\path[draw=drawColor,line width= 0.4pt,line join=round,line cap=round] ( 49.20,114.47) -- ( 43.20,114.47);

\path[draw=drawColor,line width= 0.4pt,line join=round,line cap=round] ( 49.20,138.47) -- ( 43.20,138.47);

\path[draw=drawColor,line width= 0.4pt,line join=round,line cap=round] ( 49.20,162.47) -- ( 43.20,162.47);

\path[draw=drawColor,line width= 0.4pt,line join=round,line cap=round] ( 49.20,186.47) -- ( 43.20,186.47);

\node[text=drawColor,anchor=base east,inner sep=0pt, outer sep=0pt, scale=  1.00] at ( 37.20, 63.04) {0};

\node[text=drawColor,anchor=base east,inner sep=0pt, outer sep=0pt, scale=  1.00] at ( 37.20, 87.03) {0.2};

\node[text=drawColor,anchor=base east,inner sep=0pt, outer sep=0pt, scale=  1.00] at ( 37.20,111.03) {0.4};

\node[text=drawColor,anchor=base east,inner sep=0pt, outer sep=0pt, scale=  1.00] at ( 37.20,135.03) {0.6};

\node[text=drawColor,anchor=base east,inner sep=0pt, outer sep=0pt, scale=  1.00] at ( 37.20,159.03) {0.8};

\node[text=drawColor,anchor=base east,inner sep=0pt, outer sep=0pt, scale=  1.00] at ( 37.20,183.02) {1};
\end{scope}
\begin{scope}
\path[clip] ( 49.20, 61.20) rectangle (336.15,203.75);
\definecolor{drawColor}{RGB}{169,169,169}

\path[draw=drawColor,line width= 0.4pt,dash pattern=on 4pt off 4pt ,line join=round,line cap=round] ( 49.20,180.47) -- (336.15,180.47);

\path[draw=drawColor,line width= 0.4pt,dash pattern=on 4pt off 4pt ,line join=round,line cap=round] (200.00, 61.20) -- (200.00,203.75);
\definecolor{drawColor}{gray}{0.40}

\path[draw=drawColor,line width= 0.8pt,line join=round,line cap=round] (267.20,169.67) -- (283.40,169.67);
\definecolor{drawColor}{RGB}{66,134,244}

\path[draw=drawColor,line width= 0.8pt,line join=round,line cap=round] (267.20,158.87) -- (283.40,158.87);
\definecolor{drawColor}{RGB}{102,166,30}

\path[draw=drawColor,line width= 0.8pt,line join=round,line cap=round] (267.20,148.07) -- (283.40,148.07);
\definecolor{drawColor}{RGB}{166,118,29}

\path[draw=drawColor,line width= 0.8pt,line join=round,line cap=round] (267.20,137.27) -- (283.40,137.27);
\definecolor{drawColor}{RGB}{0,0,0}

\node[text=drawColor,anchor=base west,inner sep=0pt, outer sep=0pt, scale=  0.90] at (291.50,166.57) {TC-DM};

\node[text=drawColor,anchor=base west,inner sep=0pt, outer sep=0pt, scale=  0.90] at (291.50,155.77) {Naive};

\node[text=drawColor,anchor=base west,inner sep=0pt, outer sep=0pt, scale=  0.90] at (291.50,144.97) {US};

\node[text=drawColor,anchor=base west,inner sep=0pt, outer sep=0pt, scale=  0.90] at (291.50,134.17) {RBC};
\end{scope}
\end{tikzpicture}}
			\end{minipage}\bigskip

			\caption{Simulated coverage rates of various values of parameter values and for different types of confidence sets. Based on the Setting~1 and a continuous running variable as described in the main text. Bias aware approach with true constants (TC (ref); as reference function in all graphs),  two times true constants (TC$\times2$), 0.5 times true constants (TC$\times.5$), and with rule of thumb smoothness bounds (ROT1) and (ROT2); robust bias correction (RBC); naive approach that ignores bias (Naive); and undersmoothing  (US).}\label{figurepower}
		\end{figure}
	\end{landscape}	
}  

To show that our bias-aware AR CSs not only have good coverage properties but also yield comparatively powerful inference, we simulate the rates at which the various CSs we consider cover parameter values other than the true one. We report the results for one exemplary setting (low curvature of outcome and treatment CEF, continuously distributed running variable) in Figure~\ref{figurepower}.\footnote{We focus on this setting because the coverage of the true parameter is reasonably close to the nominal level for all procedures, and thus a comparison of coverage rates at ``non-true" parameter values is meaningful across CSs. Analogous plots for other DGPs are available from the authors.} To avoid having all 16 coverage curves in one plot, we split the results into four panels: the five bias-aware AR CSs in (a), the three other AR CSs in (b), the five bias-aware DM CIs in (c), and the three other DM CIs in (d). Panels (b)--(d) also show the curve for our bias-aware AR CS with the true constants to have a common point of reference.

Panel (a) then shows that the coverage rate of bias-aware AR CSs drops very quickly to zero away from the true parameter. Panels (b)--(d) show that the coverage of bias-aware AR CSs away from the true parameter is also below that of most competing procedures over almost all the parameter space, with the exception of those which exhibit a meaningful distortion at the true parameter value and are therefore not suitable for a direct comparison. This confirms that the accurate coverage of our CSs in settings with discrete running variables and weak identification does not come at the expense of statistical power in a canonical setup, for which most competing CSs are specifically constructed.

\appendix
\section{Proofs of Theorems~1--3}\label{section:proofsofmainresults}
In this Appendix, we prove Theorems~1--3. See Online Appendix~\ref{proof:consistentcyofse} for a proof of Theorem~4. We note that, by basic least squares algebra, the statistic $\widehat\tau_{M}(h,c)$ can be written as 
\begin{align*}
&\widehat\tau_M(h,c) =\sum_{i=1}^n  w_{i}(h) M_i(c),   \quad  w_i(h)=w_{i,+}(h)-w_{i,-}(h),\\
&w_{i,+}(h) = e_1^\top Q^{-1}_+ \widetilde{X}_iK(X_i/h) \1{X_i\geq 0}, \quad Q_+ = \sum_{i=1}^n K(X_i/h) \widetilde{X}_i \widetilde{X}_i^\top \1{X_i\geq 0} \\
&w_{i,-}(h) = e_1^\top Q^{-1}_- \widetilde{X}_iK(X_i/h) \1{X_i< 0}, \quad Q_- = \sum_{i=1}^n K(X_i/h)  \widetilde{X}_i\widetilde{X}_i^\top \1{X_i< 0},
\end{align*}
where $\widetilde{X}_i=(1,X_i)^\top$.
We write $A_n(\mu) = o_{P,\mathcal{F}}(1)$ if $ \sup_{\mu \in \mathcal{F}} P(|A_n(\mu)|>\epsilon) = o(1)$ for all $\epsilon>0$ and a generic sequence $A_n(\mu)$ of random variables indexed by $\mu \in \mathcal{F}$. We also drop the dependency on $c$ from the notation for the optimal bandwidth in most instances,
and thus write $\hopt$ instead of $\hopt(c)$.

\subsection{Proof of Theorem 1}

We begin by establishing the following lemma.

\begin{lemma}\label{theorem_assumptions} Suppose that Assumption~\ref{reg1}--\ref{reg2} and either Assumption~\ref{assumptiondiscrete}  or Assumption~\ref{assumptioncontinuous} are satisfied. Then the following holds uniformly over $(\mu_Y,\mu_T)\in\mathcal{F}$: (i) $w_\textnormal{ratio}(h_{M}(c))=o_P(1)$; (ii) $(\widehat\tau_M(\widehat h_{M}(c),c) - \widehat\tau_M( h_{M}(c),c))/s_{M}( h_{M}(c),c)= o_P(1)$; and (iii) $(\bar{b}_M(\widehat h_{M}(c),c) - \bar{b}_M( h_{M}(c),c))/s_{M}( h_{M}(c),c)= o_P(1)$. 
\end{lemma}

\textit{Proof.} First suppose that Assumption~\ref{assumptiondiscrete} is satisfied, and note that  it is clear with a discrete running variable that the optimal bandwidth $\h$
shrinks with the sample size, but that it cannot tend to zero   as it has to be greater than the
support point second closest to the cutoff in order for the local linear regression estimator to be well-defined.
To  show part~(i), note that $w_\textnormal{ratio}(h_{M})$ is well-defined with probability approaching 1, and that
$$ w_\textnormal{ratio}(h_{M})\leq  \max_{i \in \{1,\dots, n\}}  \frac{w_i(\h)^2}{\sum_{j: X_j= X_i} w_j(\h)^2  }  =  \max_{i \in \{1,\dots, n\}}  \frac{1}{\sum_{j: X_j= X_i} \1{X_i=X_j} } =o_p(1)$$			
as the number of units
whose realization of the running variable is equal to any particular value in its support tends to infinity.
To show parts (ii) and (iii), recall that $\h$ approaches some value between second and third support point closest to the cutoff, and that indeed
any bandwidth  $h$  between the second and third support point closest to the cutoff  implies the same local linear
regression weights $w_i(h)$ for all $i$. Part~(ii)--(iii) then follow trivially, as each expression under consideration depends on $h$ only through $w_i(h)$.

Now suppose that Assumption~\ref{assumptioncontinuous} holds. Note that  the minimizer of the function $h\mapsto \textnormal{cv}_{1-\alpha}( r_{M}(h,c)) s_{M}(h,c)$ must balance the asymptotic bias and standard deviation of the 
the local linear regression estimator, and thus  $h_{M} \propto n^{-1/5}(1+o(1))$.
Statement~(i) then follows from classical arguments for this bandwidth. On the other hand, 
statements~(ii)--(iii)
of Lemma~\ref{theorem_assumptions} follow by applying the arguments starting in the second paragraph of the proof of Theorem E.1 in \citet{armstrong2018simple} conditional on $\mathcal{X}_n$, and the assumption that the running variable density is continuous around the cutoff. \qed

\bigskip

We now proceed with the proof of the core statement of Theorem~1.
As $\theta\in\mathcal{C}^{\alpha}_{\textnormal{ar}}$ if and only if $\tau_M(\theta)\in\mathcal{C}^{\alpha} (\theta)$, it suffices
to show that for any $c \in\IR$
\begin{align*}
	\liminf_{n\to\infty} \inf_{(\mu_Y,\mu_T)\in\mathcal{F}}	\IP(\tau_M(c) \in  \mathcal{C}^{\alpha} (c)) & \geq  1-\alpha.
\end{align*}
Note that  it follows from Lemma~\ref{theorem_assumptions}~(ii)--(iii) and uniform continuity of $\cv(\cdot)$  that
\begin{align*}
	&\frac{ |\widehat{\tau}_M( \hhopt,c ) - \tau_M(c)|}{\widehat{s}_M( \hhopt,c )} - \cv (\widehat{r}_M( \hhopt,c ))\\
	&\quad=\left| \frac{\widehat{\tau}_M(  \hopt,c  )-\E\left[\widehat{\tau}_M(  \hopt,c ) | \Xn \right]  }{ s_M(  \hopt,c )}   +\frac{b_M(  \hopt,c) }{ s_M(  \hopt,c )} \right| - \cv  (r_M(  \hopt,c ))+o_{P, \mathcal{F}}(1).
\end{align*}
By Lemma~\ref{theorem_assumptions}~(i) and Lyapunov's CLT $(\widehat{\tau}_M(  \hopt,c  )-\E\left[\widehat{\tau}_M(  \hopt,c ) | \Xn \right]) / s_M(  \hopt,c )$ converges in distribution to a standard normally distributed random variable, uniformly
over $(\mu_Y,\mu_T)\in\mathcal{F}$. It then follows again from uniform continuity of $\cv(\cdot)$ that
\begin{align*}
		\IP(\tau_M(c) \in  \mathcal{C}^{\alpha} (c))= \mathbb{P} \left( \left| S +  \frac{b_M(  \hopt,c ) }{s_M(  \hopt,c )} \right|  \leq \cv ( r_{M}( \hopt,c  ))\right) + o_{P,\mathcal{F}}(1),
\end{align*}
with $S$  a generic standard normal random variable. Honesty now follows from the definition of the critical value function $\cv(\cdot)$ if
$$\sup_{(\mu_Y,\mu_T)\in\mathcal{F}} |b_M(  \hopt,c ) / s_M(  \hopt,c )| \leq r_{M}( \hopt,c ).$$
\citet[Theorem B.3]{armstrong2018simple} implies that the last statement would hold with equality if $\mu_Y$ and $\mu_T$
had unbounded domain. We obtain a weak inequality because in our setup $\mu_T$ is naturally constrained to take values in $[0,1]$,
and  the supremum is thus taken over a smaller set of functions. This completes our proof.   \qed

\subsubsection{Alternative assumptions for discrete case}\label{app:alternative} The above asymptotic framework for the case of a discrete running variable  implies that in large samples the optimal bandwidth $h_M$ is the corner solution of the corresponding optimization problem, and is thus such that only the two closest support points on each side of the cutoff receive positive kernel weights. This in turn implies that
the corresponding  bias asymptotically dominates the corresponding standard deviation. Here we show that our CSs also remain honest, in the sense of 	\eqref{honest}, under an alternative asymptotic sequence under which the variance of $M_i(c)$ increases with the sample size at an appropriate rate. This rate is chosen such that bias and standard error are of the same stochastic  order in large samples  under the resulting optimal bandwidth.

\begin{proposition}Suppose that for each $n$, the data $\{(Y_i,T_i,X_i,),i=1,\ldots,n\}$ are i.i.d., distributed according to a law $P_n$. Under each $P_n$, the support of $X_i$ consists of the $J = J_+ + J_-$ fixed points $x_1< \ldots < x_{J_-}  <0  \leq x_{J_- +1}<\ldots <x_J$; and the conditional expectation functions $(\mu_Y,\mu_T)$ do not vary with $n$ as well. Moreover, 
	$\V(M_i(c)|X=x_j) = n  \bar\sigma^2_{M,j}(c) $ for constants  $\bar\sigma^2_{M,j}(c)>0$ for  every $c\in\IR$, $j=1,\ldots,J$, and $(\mu_Y,\mu_T)\in\mathcal{F}$; and 
	$\E( (M_i(c) - \E(M_i(c)|X_i))^q| X_i=x_j) = O(n^{q/2})$ for some $q>2$.
	Also, the kernel $K$ is a continuous, unimodal, symmetric density
	function that is equal to zero outside some compact set, say $[-1,1]$.
	 Then $\mathcal{C}_{\textnormal{ar}}^{\alpha}$
	is honest with respect to $\mathcal{F}$ in the sense of 	\eqref{honest}.
\end{proposition}

\begin{proof}We show that the statements (i)--(iii) of Lemma A.1 also hold under the conditions of the Proposition. Honesty of our CS then follows as in the proof of Theorem~1 by noting that  Lyapunov's CLT still applies under the conditions of the Proposition.
	
	 First note that for any fixed bandwidth $h >\max\{|x_{J_- -1}|,x_{J_- +2} \}$, we have that both $\bar{b}_M( h,c)$ and $s_{M}(h,c)$ converge with rate $n^{-1/2}$ to strictly positive constants in probability, and thus 	$h_M = \bar{h}_M + o_P(1)$, with $\bar{h}_M > \max\{|x_{J_- -1}|,x_{J_- +2} \}$.
	  Statement~(i) then follows from the same reasoning as in the proof of Lemma~A.1(i).  To show statement~(ii), note that because $s_{M}( h_{M},c) = O_P(1)$,  it suffices to show that $\widehat\tau_M(\widehat h_{M},c) - \widehat\tau_M( h_{M},c)$ converges to zero in probability. With a discrete running variable, we can write the estimator $\widehat\tau_M(h,c)$ as follows:
\begin{align*}
	&\widehat\tau_M(h,c)= \sum_{j=1}^J   w_{(j)}(h)\bar{M}_j(c), \quad\bar{M}_j(c) = \frac{1}{n \widehat p_j}\sum_{i:X_i=x_j}M_i(c) ,\quad  w_{(j)}(h)=w_{(j)+}(h)-w_{(j)-}(h),\\
	&w_{(j)+}(h) = e_1^\top \tilde Q^{-1}_+   \widehat p_jK(x_j/h) \widetilde{x}_j  \1{x_j\geq 0}, \quad\tilde Q_+ = \sum_{j=1}^J  \widehat p_j K(x_j/h) \widetilde{x}_j \widetilde{x}_j^\top \1{x_j\geq 0} \\
	&w_{(j)-}(h) =  e_1^\top \tilde Q^{-1}_-  \widehat p_jK(x_j/h) \widetilde{x}_j  \1{x_j< 0}, \quad \tilde Q_- = \sum_{j=1}^J  \widehat p_j  K(x_j/h)  \widetilde{x}_j\widetilde{x}_j^\top \1{x_j< 0},
\end{align*}
with  $\widetilde{x}_j =(1,x_j)^\top$ and  $\widehat p_j =\sum_{i=1}^n \1{X_i=x_j}/n$ the relative frequency of the $j$th support point in the data. We then have
\begin{align*}
	\widehat\tau_M(\widehat h_{M},c) - \widehat\tau_M( h_{M},c) &= 
	\sum_{j=1}^J  ( w_{(j)}(\widehat h_{M})- w_{(j)}( h_{M})) \mu_M(x_j,c) \nonumber \\ &\quad +
	\sum_{j=1}^J  ( w_{(j)}(\widehat h_{M})- w_{(j)}( h_{M})) (\bar{M}_j(c)-\mu_M(x_j,c)).
\end{align*}
From continuity of the kernel function, it follows that the mapping $h\mapsto w_{(j)}(h)$ is continuous for all $j=1,\ldots,J$, and hence $w_{(j)}(\widehat h_{M})- w_{(j)}( h_{M})=o_P(1)$ for all $j=1,\ldots,J$.
The first term on the right-hand side of the previous equation then converges to zero in probability follows because $\mu_M(\cdot,c)$ has uniformly bounded second derivatives, and from the algebraic properties of the local linear regression weights. For the second term, convergence in probability to zero follows because $\bar{M}_j(c)-\mu_M(x_j,c) =O_P(1)$. Taken together, this proves statement~(ii). Statement~(iii) can then be shown analogously.
\end{proof}

\subsection{Proof of Theorem \ref{theorem_formofCS}} 
To simplify the exposition, we emphasize the dependence of various estimators on $c$ in our
notation, but suppress  their dependency on the bandwidth~$h$ (which does not depend
on $c$ under the conditions of this theorem). The CS $ \mathcal{C}_{\textnormal{ar}}^{\alpha}(h)$ is given by the set of all values of $c$ satisfying
$$\vartheta(c) \leq 0, \quad\textnormal{ where }\quad \vartheta(c)  \equiv |\htauY - c \htauT| - \textnormal{cv}_{1-\alpha}(\widehat{r}_M(c))  \widehat s_M(c). $$
The function $\vartheta(c)$ is  continuous because   $\textnormal{cv}_{1-\alpha}$ is uniformly continuous, and both the standard error $\widehat s_M(c)=(\widehat s_Y^2 - 2c \widehat s_{TY} +c^2 \widehat s_T^2)^{1/2}$  and the worst case bias
$\overline{b}_{M}(h,c)=  - (B_Y + |c|  B_T)/2\cdot \sum_{i=1}^n w_{i}(h) X_i^2\sgn(X_i)$ are continuous  in $c$. 
The term $\textnormal{cv}_{1-\alpha}(\widehat{r}_M(c))  \widehat s_M(c) $ is  also 
strictly convex in $c$,  because  both the standard error and the worst-case bias are  convex in $c$  and
$\textnormal{cv}_{1-\alpha}(\cdot)$ is strictly convex and increasing. 
The shape of $ \mathcal{C}_{\textnormal{ar}}^{\alpha}(h)$ is then determined by the roots of $\vartheta(c)$. 
 To prove the theorem, it suffices to
show that the function $\vartheta(c)$ always fits into one of the following four
categories:  (i) $\vartheta(c)\leq 0$ for all $c$; (ii) $\vartheta(c)$ has two roots, and there exists  $c^*>0$ such that $\vartheta(c) < 0 $ for all $|c|>c^*$; (iii)  $\vartheta(c)$ has two roots, and there exists $c^*>0$ such that  $\vartheta(c) > 0$ for all $|c|>c^*$, and (iv)  $\vartheta(c)$  has one root.
Then $ \mathcal{C}_{\textnormal{ar}}^{\alpha}(h)= \IR$ in case (i),  $ \mathcal{C}_{\textnormal{ar}}^{\alpha}(h)=  (-\infty, a_1] \cup [a_2, \infty)$ 
for some $a_1<a_2$ in case (ii); and by $\mathcal{C}_{\textnormal{ar}}^{\alpha}(h)= [a_1, a_2]$  for some $a_1<a_2$  in case (iii), and $\mathcal{C}_{\textnormal{ar}}^{\alpha}(h)= (-\infty, a_2]$ or  $\mathcal{C}_{\textnormal{ar}}^{\alpha}(h)= [a_1, \infty)$ in case (iv). We now go through a
number of case distinctions.

If  $\widehat \tau_T = 0$, then $|\htauY-c\htauT|$ is a constant function in $c$. As  $\textnormal{cv}_{1-\alpha}(\widehat{r}_M(c))    \widehat s_M(c)$  is strictly  convex  in $c$ and unbounded, $\vartheta(c)$ must be  either of form (i) or (ii). We threfore suppose
that $\htauT \neq 0$ from now on, and write $\widehat \poi = \htauY / \htauT$ .  As $\vartheta(\widehat \poi )<0$ by construction, the function
$\vartheta(c)$ cannot be strictly positive.
As $|\htauY-c\htauT|$ is a piecewise linear function and $\textnormal{cv}_{1-\alpha}(\widehat{r}_M(c))   \widehat s_M(c)$  is strictly convex, the function $\vartheta(c)$ can also have at most two roots for $c \leq \widehat \poi $, and at most two roots for $c >\widehat \poi  $. If it does not have any root, $\vartheta(c)$ is of the form (i).

Let us first assume that $\lim_{c \to \pm \infty } \vartheta(c)\neq 0$.	It follows from basic algebra that there exists some $c^*$ sufficiently large such that
$ \sgn(\vartheta(c))  =  \sgn( \vartheta(-c)) =1 $ or  $ \sgn(\vartheta(c))  =  \sgn( \vartheta(-c)) =-1 $    and  $ \vartheta(c) \neq 0$ for all $c > c^*$.	 The function	$\vartheta(c)$  therefore cannot have one or three roots; so it must have either four roots or two roots or none. If  $ \sgn(\vartheta(c))   =-1$ for all $|c| > c^*$,  which means that  $|\htauY - c \htauT| > \textnormal{cv}_{1-\alpha}(\widehat{r}_M(c))   \widehat s_M(c)$. The function $\textnormal{cv}_{1-\alpha}(\widehat{r}_M(c))    \widehat s_M(c)$  intersects once with the function $|\htauY-c\htauT|$ for $c < \widehat \poi $, and once for $c >\widehat \poi $. Therefore $\vartheta(c)$ must be of form (iii) in this case. If  $ \sgn(\vartheta(c))   =-1$ for all $|c| > c^*$, the 
above reasoning only yields that $\vartheta(c)$ has at most four roots. 
However,  note  that  for $|c| \rightarrow \infty$ the absolute value of the first derivative of $\textnormal{cv}_{1-\alpha}(\widehat{r}_M(c)) \widehat s_M(c)$ with respect to $c$ converges to some constant $\varpi$, and that for any value of $\varsigma \in \mathbb{R} $ the expression $\sgn(c) \cdot ( \textnormal{cv}_{1-\alpha}(\widehat{r}_M(c)) \widehat s_M - |\varsigma +\varpi \cdot c |)  $ converges to a constant. Choose $\varsigma$ such that the latter constant is zero, and set $\varrho(c) =|\varsigma +\varpi c | $.
By construction, $\varrho(c)$ intersects  with    $|\htauY- c \htauT|$   twice either for   $c \leq \widehat \poi $ or  $c \geq \widehat \poi $.
It also holds that $\varrho (c) \leq \textnormal{cv}_{1-\alpha}(\widehat{r}(c))   \widehat s_M(c)$ for all $c$ by strict convexity of $\textnormal{cv}_{1-\alpha}(\widehat{r}(c))   \widehat s_M(c)$. This reasoning implies that $\vartheta(c)$ can have at most two roots, and must be of form (ii) in this case.

Now suppose  that $\lim_{c \to \pm \infty } \vartheta(c)= 0$, which 
only occurs if $\widehat\tau_T = \pm \textnormal{cv}_{1-\alpha}(\widehat{r}_T(c)) \cdot \widehat s_T(c)$. It   then follows from strict convexity of $\textnormal{cv}_{1-\alpha}(\widehat{r}_M(c))  \widehat s_M(c)$ that $\vartheta(c)$ cannot have three roots. $\vartheta(c)$  is therefore of form (i)  if it does not have any root, and otherwise of form (iv).
This  completes the proof.    \qed

\subsection{Proof of Theorem \ref{theorem_delta_ar}} We begin by giving a formal description
of a bias-aware DM CI. Recall the definition of
$U_i$ from Section~\ref{subsection:frdinference}, and let $b_{U}(h) = \E(\widehat\tau_U(h)|\mathcal{X}_n)$ and
$s_{U}(h)=\V(\widehat\tau_U(h)|\mathcal{X}_n)^{1/2}$ denote 
conditional bias and standard deviation, respectively, of the SRD-type
estimator $\widehat\tau_U(h)$.
Exploiting linearity, one can write
\begin{align*}
b_{U}(h) =\sum_{i=1}^n  w_{i}(h) (\mu_U(X_i) - \tau_{U}) \textnormal{ and }s_{U}(h) = \left(\sum_{i=1}^n  w_i(h)^2 \sigma_{U,i}^2\right)^{1/2},
\end{align*}
where $\mu_U(x)=(\mu_Y(x) -\tau_Y)/\tau_T- \tau_Y(\mu_T(x)-\tau_T)/\tau_T^2$ depends on the functions $\mu_Y$ and $\mu_T$, and $\sigma^2_{U,i} = \V(U_i|X_i)$ is the conditional variance of $U_i$ given $X_i$.
Because the bias depends on  $(\mu_Y,\mu_T)$ through
the function $\mu_U \in \mathcal{F}_H(B_Y/|\tau_T| + |\tau_Y| B_T/\tau_T^2)$ only, its ``worst case'' magnitude over the functions contained in $\mathcal{F}^\delta$ is 
\begin{align*}
\sup_{(\mu_Y,\mu_T)\in\mathcal{F}^\delta}| b_{U}(h)|= \overline{b}_U (h) \equiv - \frac{1}{2}\left(\frac{B_Y}{|\tau_T|}+\frac{|\tau_Y| B_T}{\tau_T^2}\right)\sum_{i=1}^n  w_{i}(h)X_i^2  \sgn(X_i).
\end{align*}
An infeasible bias-aware DM CI is then given by
\begin{align}
\mathcal{C}_{\Delta}^{\alpha} = \left[\hpoi(h_U)\pm \textnormal{cv}_{1-\alpha}\left(  \overline{b}_U (h_U)/ s_U (h_U)\right) s_U (h_U)\right], \label{c_delta}
\end{align}
where $ h_U = \argmin_h \textnormal{cv}_{1-\alpha}\left( {\overline{b}}_U (h)/  s_U (h)\right)  s_U (h)$  is the bandwidth that minimizes its length. It is easy to see that this optimal bandwidth must such that neither bias nor variance dominate asymptotically, which means that $ \lim n^{-1/5} h_U =c > 0$.

Making this CI feasible would require three main modifications: (i) replacing the unknown bias bound with an estimate
$\widehat{\overline{b}}_U (h)$ which replaces $\tau_Y$ and $\tau_T$ with feasible estimates, such as local
linear estimates  $\widehat\tau_Y = \widehat\tau_Y(g_Y)$ and $\widehat\tau_T = \widehat\tau_T(g_T)$
based on  preliminary bandwidths $g_Y$ and $g_T$; (ii) replacing the standard deviation $s_U(h)$
with a valid standard error, which could be achieved as in Section \ref{subsectionstderrors}
using estimates of the form $\widehat U_i =  (Y_i -\widehat\tau_Y)/\widehat\tau_T- \widehat\tau_Y(T_i-\widehat\tau_T)/\widehat\tau_T^2$ of the $U_i$); (iii)
replacing the bandwidth $h_U$  with an empirical analogue, like an adaptation of the  procedure in Section~\ref{sec:bandwidth_choice}. As such modifications can be shown not to affect the first-order
asymptotic coverage properties of the CI under standard regularity conditions, we  base our result on a comparison of $\mathcal{C}_{*}^{\alpha}$ and $\mathcal{C}_{\Delta}^{\alpha}$. 

To prove Theorem \ref{theorem_delta_ar}, we  make the dependence of quantities like
$h_M(c)$ on $c$ again explicit. We begin by noting that the events
$\poi^{(n)}\in  \mathcal{C}_{\Delta}^{\alpha}$ and $\poi^{(n)}\in  \mathcal{C}_{*}^{\alpha}$ occur if and only if
\begin{align}
&\frac{|\hpoi( h_U) -\poi^{(n)}|}{  s_U ( h_U)}-\textnormal{cv}_{1-\alpha}\left(\frac{ \overline{b}_U ( h_U)}{  s_U ( h_U)}\right)\leq 0 \label{Th3Eq1}\\
\textnormal{and}\quad&\frac{|\widehat\tau_M(  h_M(\poi^{(n)}),\poi^{(n)})|}{ s_M (  h_M(\poi^{(n)}),\poi^{(n)})}-\textnormal{cv}_{1-\alpha}\left(\frac{\overline{b}_M (  h_M(\poi^{(n)}),\poi^{(n)})}{  s_M (  h_M(\poi^{(n)}),\poi^{(n)})}\right)\leq 0,\label{Th3Eq2}
\end{align}
respectively. Because the left-hand sides of the last two displays  both approximately behave like a constant plus the absolute value of a normal random variable with variance 1 in large samples, it  suffices to show that
the difference between the respective left-hand sides of the last two
displays  converges to zero in probability, uniformly over $\mathcal{F}^\delta$. To
show this, note first that standard delta method arguments yield that the left-hand side
of~\eqref{Th3Eq1} is equal to
$$\frac{|\widehat\tau_U(h_U) -\kappa n^{-2/5}|}{ s_U (h_U)}-\textnormal{cv}_{1-\alpha}\left(\frac{\overline{b}_U (h_U)}{ s_U (h_U)}\right) + o_{P,\mathcal{F}^{\delta}}(1).$$
Next, note that $U_i = M_i(\theta)/\tau_T$, and that we thus have that
$$\widehat\tau_U(h) = \frac{\widehat\tau_M(h,\poi)}{\tau_T},\qquad 
s_U (h) = \frac{s_M (h,\poi)}{|\tau_T|}, \qquad \overline{b}_U(h) = \frac{\overline{b}_M(h,\poi)}{|\tau_T|}, $$
for any $h>0$. Substituting these identities into the definition of $h_U$, we also find that
$$ h_U = \argmin_h \textnormal{cv}_{1-\alpha}\left(\frac{{\overline{b}}_M (h,\poi)}{s_M (h,\poi)}\right)\cdot \frac{s_M (h,\poi)}{|\tau_T|} = \argmin_h \textnormal{cv}_{1-\alpha}\left(\frac{{\overline{b}}_M (h,\poi)}{s_M (h,\poi)}\right)  s_M (h,\poi)  = h_M(\poi).$$
The left-hand side of~\eqref{Th3Eq1} is thus equal to
\begin{align*}
\frac{|\widehat\tau_M(h_M(\poi),\poi) -\tau_T\kappa n^{-2/5}|}{ s_M (h_M(\poi),\poi)}-\textnormal{cv}_{1-\alpha}\left(\frac{\overline{b}_M (h_M(\poi),\poi)}{s_M (h_M(\poi),\poi)}\right)+o_{P,\mathcal{F}^{\delta}}(1).
\end{align*}
Now consider the term on the left-hand side of~\eqref{Th3Eq2}. By simple algebra, we have that for $n$ sufficiently large 
\begin{align*}
\overline{b}_M (h,\poi^{(n)})  &= \overline{b}_M (h,\poi) + ( |\theta + n^{-2/5} \kappa|  - |\theta| ) \overline{b}_T (h) =   \overline{b}_M (h,\poi) +  |n^{-2/5} \kappa| \overline{b}_T (h)    ,\\
s_M (h,\poi^{(n)})^2  &= s_M^2 (h,\poi) + n^{-2/5} \kappa  \; (  (2 \theta+  n^{-2/5} \kappa)  s_T^2 (h) -2  \tilde{s}_{M(\poi),T} (h)  ),
\end{align*}
with $\tilde{s}_{M(\poi),T} (h)=(\sum_{i=1}^n  w_i(h)^2 \sigma_{M(\poi),T,i})^{1/2}$ a conditional covariance term of the same order as $s_T (h)$.
These identities imply that $\widehat\tau_M(h,\poi)$
and $\widehat\tau_M(h,\poi^{(n)})$ have the same first-order bias and standard deviation along any bandwidth sequence $h$ of order $n^{-1/5}$; and from Theorem~2.1(i) in \citet{armstrong2018simple}, we know that to first order the optimal bandwidth that minimizes the length of a bias-aware CI depends on the first-order bias and standard deviation only. This yields that $h_M(\poi^{(n)}) = h_M(\poi)(1+o_{P,\mathcal{F}^{\delta}}(1))$.
Arguing as in the proof of Lemma~\ref{theorem_assumptions}, the left-hand side of~\eqref{Th3Eq2}  is thus equal to 
\begin{align*}
	\frac{|\widehat\tau_M(h_M(\poi),\poi) -\tau_T\kappa n^{-2/5}|}{ s_M (h_M(\poi),\poi)}-\textnormal{cv}_{1-\alpha}\left(\frac{\overline{b}_M (h_M(\poi),\poi)}{s_M (h_M(\poi),\poi)}\right)+o_{P,\mathcal{F}^{\delta}}(1),
\end{align*}
which completes the proof.\qed

\nocite{van1996weak} 
\bibliography{bibl}  

\begin{thebibliography}{24}
\newcommand{\enquote}[1]{``#1''}
\expandafter\ifx\csname natexlab\endcsname\relax\def\natexlab#1{#1}\fi

\bibitem[\protect\citeauthoryear{Abadie and Imbens}{Abadie and
  Imbens}{2006}]{abadie2006large}
\textsc{Abadie, A. and G.~W. Imbens} (2006): \enquote{Large Sample Properties
  of Matching Estimators for Average Treatment Effects,} \emph{Econometrica},
  74, 235--267.

\bibitem[\protect\citeauthoryear{Abadie, Imbens, and Zheng}{Abadie
  et~al.}{2014}]{abadie2014inference}
\textsc{Abadie, A., G.~W. Imbens, and F.~Zheng} (2014): \enquote{Inference for
  misspecified models with fixed regressors,} \emph{Journal of the American
  Statistical Association}, 109, 1601--1614.

\bibitem[\protect\citeauthoryear{Anderson and Rubin}{Anderson and
  Rubin}{1949}]{anderson1949estimation}
\textsc{Anderson, T. and H.~Rubin} (1949): \enquote{Estimation of the
  parameters of a single equation in a complete system of stochastic
  equations,} \emph{Annals of Mathematical Statistics}, 20, 46--63.

\bibitem[\protect\citeauthoryear{Andrews, Stock, and Sun}{Andrews
  et~al.}{2019}]{andrews2019weak}
\textsc{Andrews, I., J.~H. Stock, and L.~Sun} (2019): \enquote{Weak Instruments
  in Instrumental Variables Regression: Theory and Practice,} \emph{Annual
  Review of Economics}, 11, 727--753.

\bibitem[\protect\citeauthoryear{Armstrong and Koles{\'a}r}{Armstrong and
  Koles{\'a}r}{2018}]{armstrong2018optimal}
\textsc{Armstrong, T. and M.~Koles{\'a}r} (2018): \enquote{Optimal inference in
  a class of regression models,} \emph{Econometrica}, 86, 655--683.

\bibitem[\protect\citeauthoryear{Armstrong and Koles{\'a}r}{Armstrong and
  Koles{\'a}r}{2020}]{armstrong2018simple}
---\hspace{-.1pt}---\hspace{-.1pt}--- (2020): \enquote{Simple and honest
  confidence intervals in nonparametric regression,} \emph{Quantitative
  Economics}.

\bibitem[\protect\citeauthoryear{Armstrong and Koles{\'a}r}{Armstrong and
  Koles{\'a}r}{2021}]{armstrong2018finite}
---\hspace{-.1pt}---\hspace{-.1pt}--- (2021): \enquote{Finite-sample optimal
  estimation and inference on average treatment effects under
  unconfoundedness,} \emph{Econometrica}, 89, 1141--1177.

\bibitem[\protect\citeauthoryear{Battistin, Brugiavini, Rettore, and
  Weber}{Battistin et~al.}{2009}]{battistin2009retirement}
\textsc{Battistin, E., A.~Brugiavini, E.~Rettore, and G.~Weber} (2009):
  \enquote{The retirement consumption puzzle: evidence from a regression
  discontinuity approach,} \emph{American Economic Review}, 99, 2209--26.

\bibitem[\protect\citeauthoryear{Bertanha and Moreira}{Bertanha and
  Moreira}{2018}]{bertanha2016impossible}
\textsc{Bertanha, M. and M.~J. Moreira} (2018): \enquote{Impossible Inference
  in Econometrics: Theory and Applications,} \emph{Journal of Econometrics}.

\bibitem[\protect\citeauthoryear{Calonico, Cattaneo, and Titiunik}{Calonico
  et~al.}{2014}]{calonico2014robust}
\textsc{Calonico, S., M.~D. Cattaneo, and R.~Titiunik} (2014): \enquote{Robust
  nonparametric confidence intervals for regression-discontinuity designs,}
  \emph{Econometrica}, 82, 2295--2326.

\bibitem[\protect\citeauthoryear{Card, Lee, Pei, and Weber}{Card
  et~al.}{2015}]{card2015inference}
\textsc{Card, D., D.~S. Lee, Z.~Pei, and A.~Weber} (2015): \enquote{Inference
  on causal effects in a generalized regression kink design,}
  \emph{Econometrica}, 83, 2453--2483.

\bibitem[\protect\citeauthoryear{Fan and Gijbels}{Fan and
  Gijbels}{1996}]{fan1996local}
\textsc{Fan, J. and I.~Gijbels} (1996): \emph{Local polynomial modelling and
  its applications}, Chapman \& Hall/CRC.

\bibitem[\protect\citeauthoryear{Feir, Lemieux, and Marmer}{Feir
  et~al.}{2016}]{feir2016weak}
\textsc{Feir, D., T.~Lemieux, and V.~Marmer} (2016): \enquote{Weak
  identification in fuzzy regression discontinuity designs,} \emph{Journal of
  Business \& Economic Statistics}, 34, 185--196.

\bibitem[\protect\citeauthoryear{Hahn, Todd, and Van~der Klaauw}{Hahn
  et~al.}{2001}]{hahn2001identification}
\textsc{Hahn, J., P.~Todd, and W.~Van~der Klaauw} (2001):
  \enquote{Identification and Estimation of Treatment Effects with a
  Regression-Discontinuity Design,} \emph{Econometrica}, 69, 201--209.

\bibitem[\protect\citeauthoryear{Imbens and Kalyanaraman}{Imbens and
  Kalyanaraman}{2012}]{imbens2012optimal}
\textsc{Imbens, G. and K.~Kalyanaraman} (2012): \enquote{Optimal bandwidth
  choice for the regression discontinuity estimator,} \emph{Review of Economic
  Studies}, 79, 933--959.

\bibitem[\protect\citeauthoryear{Imbens and Manski}{Imbens and
  Manski}{2004}]{imbens2004confidence}
\textsc{Imbens, G. and C.~Manski} (2004): \enquote{Confidence Intervals for
  Partially Identified Parameters,} \emph{Econometrica}, 72, 1845--1857.

\bibitem[\protect\citeauthoryear{Imbens and Wager}{Imbens and
  Wager}{2019}]{imbens2019optimized}
\textsc{Imbens, G. and S.~Wager} (2019): \enquote{Optimized regression
  discontinuity designs,} \emph{Review of Economics and Statistics}, 101,
  264--278.

\bibitem[\protect\citeauthoryear{Imbens and Lemieux}{Imbens and
  Lemieux}{2008}]{imbens2008regression}
\textsc{Imbens, G.~W. and T.~Lemieux} (2008): \enquote{Regression discontinuity
  designs: A guide to practice,} \emph{Journal of Econometrics}, 142, 615--635.

\bibitem[\protect\citeauthoryear{Kamat}{Kamat}{2018}]{kamat2018nonparametric}
\textsc{Kamat, V.} (2018): \enquote{On nonparametric inference in the
  regression discontinuity design,} \emph{Econometric Theory}, 34, 694--703.

\bibitem[\protect\citeauthoryear{Kolesár and Rothe}{Kolesár and
  Rothe}{2018}]{kolesar2018discrete}
\textsc{Kolesár, M. and C.~Rothe} (2018): \enquote{Inference in Regression
  Discontinuity Designs with a Discrete Running Variable,} \emph{American
  Economic Review}, 108, 2277–--2304.

\bibitem[\protect\citeauthoryear{Li}{Li}{1989}]{li1989honest}
\textsc{Li, K.-C.} (1989): \enquote{Honest confidence regions for nonparametric
  regression,} \emph{Annals of Statistics}, 17, 1001--1008.

\bibitem[\protect\citeauthoryear{Low}{Low}{1997}]{low1997nonparametric}
\textsc{Low, M.} (1997): \enquote{On nonparametric confidence intervals,}
  \emph{Annals of Statistics}, 25, 2547--2554.

\bibitem[\protect\citeauthoryear{Staiger and Stock}{Staiger and
  Stock}{1997}]{staiger1997instrumental}
\textsc{Staiger, D. and J.~H. Stock} (1997): \enquote{Instrumental Variables
  Regression with Weak Instruments,} \emph{Econometrica}, 557--586.

\bibitem[\protect\citeauthoryear{van~der Vaart and Wellner}{van~der Vaart and
  Wellner}{1996}]{van1996weak}
\textsc{van~der Vaart, A. and J.~Wellner} (1996): \emph{Weak Convergence and
  Empirical Processes}, Springer.

\end{thebibliography}


\begin{thebibliography}{0}
\newcommand{\enquote}[1]{``#1''}
\expandafter\ifx\csname natexlab\endcsname\relax\def\natexlab#1{#1}\fi

\end{thebibliography}
\end{document}